\documentclass[twocolumn,showpacs,preprintnumbers,amsmath,amssymb,APSl,prd,nofootinbib,superscriptaddress]{revtex4-2}
\usepackage{graphicx,color}
\usepackage{amsmath}
\usepackage{amssymb}
\usepackage{hyperref}
\usepackage[utf8]{inputenc}
\usepackage[english]{babel}
\usepackage{epsfig}
\usepackage{subfigure}
\usepackage{wasysym}
\usepackage{color,xcolor}
\usepackage{amsmath}
\usepackage{bm}
\usepackage{epsfig}
\usepackage{amsfonts}
\usepackage{dcolumn}
\usepackage{float}

\hypersetup{colorlinks=true, linkcolor=blue, citecolor=green}

\usepackage{dcolumn}
\usepackage{bm}
\usepackage{ifpdf}
\usepackage{hyperref}
\usepackage{float}
\usepackage{bm}
\usepackage{xcolor,color,graphicx,graphics}
\usepackage[OT1]{fontenc}
\usepackage{latexsym,amssymb,amsmath,amsfonts}
\usepackage{makeidx}
\usepackage{epsfig}
\usepackage{epstopdf}
\usepackage{mathrsfs}
\hypersetup{colorlinks=true, linkcolor=blue, citecolor=green}
\usepackage{enumerate}
\usepackage{xcolor}
 \usepackage{multirow}

\begin{document}

\title{Free-falling test particles in a charged Kalb-Ramond black hole:\\ gravitational Doppler effect and tidal forces}

  \author{Daniela S. J. Cordeiro} 
        \email{fc52853@alunos.ciencias.ulisboa.pt}        
\affiliation{Instituto de Astrof\'{i}sica e Ci\^{e}ncias do Espa\c{c}o, Faculdade de Ci\^{e}ncias da Universidade de Lisboa, Edifício C8, Campo Grande, P-1749-016 Lisbon, Portugal}
	\author{Ednaldo L. B. Junior} \email{ednaldobarrosjr@gmail.com}
\affiliation{Faculdade de F\'{i}sica, Universidade Federal do Pará, Campus Universitário de Tucuruí, CEP: 68464-000, Tucuruí, Pará, Brazil}
\affiliation{Programa de P\'{o}s-Gradua\c{c}\~{a}o em F\'{i}sica, Universidade Federal do Sul e Sudeste do Par\'{a}, 68500-000, Marab\'{a}, Par\'{a}, Brazill}
     \author{José Tarciso S. S. Junior}
    \email{tarcisojunior17@gmail.com}
\affiliation{Faculdade de F\'{i}sica, Programa de P\'{o}s-Gradua\c{c}\~{a}o em F\'{i}sica, Universidade Federal do Par\'{a}, 66075-110, Bel\'{e}m, Par\'{a}, Brazill}
	\author{Francisco S. N. Lobo} \email{fslobo@ciencias.ulisboa.pt}
\affiliation{Instituto de Astrof\'{i}sica e Ci\^{e}ncias do Espa\c{c}o, Faculdade de Ci\^{e}ncias da Universidade de Lisboa, Edifício C8, Campo Grande, P-1749-016 Lisbon, Portugal}
\affiliation{Departamento de F\'{i}sica, Faculdade de Ci\^{e}ncias da Universidade de Lisboa, Edif\'{i}cio C8, Campo Grande, P-1749-016 Lisbon, Portugal}
    \author{\\Manuel E. Rodrigues} \email{esialg@gmail.com}
\affiliation{Faculdade de F\'{i}sica, Programa de P\'{o}s-Gradua\c{c}\~{a}o em F\'{i}sica, Universidade Federal do Par\'{a}, 66075-110, Bel\'{e}m, Par\'{a}, Brazill}
\affiliation{Faculdade de Ci\^{e}ncias Exatas e Tecnologia, Universidade Federal do Par\'{a}, Campus Universit\'{a}rio de Abaetetuba, 68440-000, Abaetetuba, Par\'{a}, Brazil}
 \author{Diego Rubiera-Garcia} \email{ drubiera@ucm.es}
\affiliation{Departamento de Física Téorica and IPARCOS, Universidad Complutense de Madrid, E-28040 Madrid, Spain}
     \author{Luís F. Dias da Silva} 
        \email{fc53497@alunos.fc.ul.pt}
\affiliation{Instituto de Astrof\'{i}sica e Ci\^{e}ncias do Espa\c{c}o, Faculdade de Ci\^{e}ncias da Universidade de Lisboa, Edifício C8, Campo Grande, P-1749-016 Lisbon, Portugal}
    \author{Henrique A. Vieira} \email{henriquefisica2017@gmail.com}
\affiliation{Faculdade de F\'{i}sica, Programa de P\'{o}s-Gradua\c{c}\~{a}o em F\'{i}sica, Universidade Federal do Par\'{a}, 66075-110, Bel\'{e}m, Par\'{a}, Brazill}

\begin{abstract}

Space-times exhibiting spontaneous Lorentz symmetry-breaking have recently attracted much attention, with Kalb-Ramond (KR) gravity providing a notable example. In this context, we examine the free-fall motion of a test particle toward an electrically charged black hole arising from the coupling of the KR field with the Maxwell one in General Relativity. We investigate how the Lorentz symmetry-breaking parameter affects the free-fall velocity of the particle as it approaches black hole inner regions. Additionally, we analyze the influence of this parameter on the emission and detection of signals by observers in different frames. We furthermore explore modifications to the radial and angular components of tidal forces in this space-time and compare the results with those obtained for the Reissner-Nordström black hole. Finally, we analytically solve the geodesic deviation equation under two different conditions, identifying a subtle effect of the Lorentz symmetry breaking parameter in the charged KR metric, and compare it with two other space-time metrics with spontaneous symmetry breaking. These findings provide useful insights into how models of spontaneous Lorentz symmetry-breaking influence gravitational dynamics in the space-times of charged black holes.

\end{abstract}

\date{\today}

\maketitle

\section{Introduction}

In 1915, Einstein introduced the Theory of General Relativity (GR), an extension of Special Relativity founded on the principles of energy conservation for isolated systems and the equivalence principle \cite{Einstein1, Einstein2}. The latter states that the acceleration experienced by a free-falling object under gravity is indistinguishable from that of an accelerated object in deep space, far from any gravitational influence \cite{Inverno}. Over the past century, this groundbreaking theory has profoundly reshaped our understanding of the universe, leading to the discovery of numerous astrophysical and cosmological phenomena. Among them, one the most enigmatic and fascinating are black holes, regions of space where gravity is so extreme that everything (even light) falling into its vicinity gets trapped. These objects challenge our deepest intuitions about space, time, and the very fabric of reality, continuing to inspire new research and discoveries in modern physics.
  
Black holes arise as exact solutions of Einstein's field equations and are understood as the ultimate outcome of the complete gravitational collapse of massive stars that have exhausted their nuclear fuel. This collapse gives rise to two fundamental features: the formation of an event horizon, which acts as a one-way membrane separating the interior and exterior regions of space-time \cite{Wald}, and the development of a singularity, characterized by the existence of incomplete geodesic paths \cite{Senovilla:2014gza}.
The first known black hole solution was discovered by Karl Schwarzschild in 1916 \cite{S}, representing a vacuum solution to Einstein’s equations. This was later generalized to include electric charge in the form of the Reissner-Nordström (RN) solution and further extended to rotating black holes with the Kerr solution \cite{K1}, which is now recognized as the most astrophysically relevant case.

The existence of black holes was  confirmed in 2015 when the Laser Interferometer Gravitational-Wave Observatory (LIGO) detected gravitational wave signals interpreted as the merger of two black holes \cite{LIGO, LIGO2}. This was further reinforced in 2019 by the Event Horizon Telescope (EHT), which captured the first direct image of a black hole’s shadow in the supermassive core of the M87 galaxy \cite{EHT1, EHT2, EHT3, EHT4, EHT5, EHT6}, followed in 2022 by the imaging of Sagittarius A* (Sgr A*), the black hole at the center of our Milky Way \cite{EHT7, EHT8, EHT9, EHT10, EHT11, EHT12, EHT13, EHT14, EHT15}. These groundbreaking observations have significantly strengthened our confidence in both the existence of black holes and the robustness of GR in describing gravitational physics across the scales tested so far.

Despite the success of GR in describing gravitational phenomena, both theoretical considerations and observational data leave room for the possibility of new gravitational physics in the strong-field regime \cite{Bull:2015stt}. This has opened a fertile area of research, exploring modifications to GR that could give rise to novel compact objects and physical effects. One such modification that has recently gained attention in the literature involves the violation of Lorentz invariance (LI), the fundamental principle that all local reference frames are equivalent in describing the laws of physics, a symmetry that GR upholds at every point in spacetime (see, e.g., \cite{Addazi:2021xuf} for a review on the current status of these theories). In LI-violating theories, this symmetry is broken at a specific energy scale, as suggested by various approaches, including string theory models \cite{corda1, corda2, corda3, corda4}, loop quantum gravity \cite{LQG1, LQG2, LQG3, LQG4}, non-commutative field theory \cite{NC, NC1, NC2}, massive gravity \cite{MG1, MG2}, and Horava-Lifshitz gravity \cite{Horava}.

Lorentz symmetry-breaking can occur in two distinct ways, each with different theoretical and phenomenological implications. In explicit breaking, the Lagrangian itself is not invariant under Lorentz transformations, meaning that the fundamental equations governing the system directly violate Lorentz symmetry at all energy scales \cite{Bianch}. This type of breaking typically introduces preferred directions or frames at the level of the action, which can lead to modified dispersion relations and potential observational signatures in high-energy physics and cosmology.
On the other hand, in spontaneous Lorentz symmetry breaking, the Lagrangian remains formally Lorentz-invariant, but the vacuum expectation value of a dynamical field selects a preferred direction in spacetime. This mechanism, analogous to spontaneous symmetry breaking in particle physics, results in emergent Lorentz-violating effects while preserving the underlying structure of the theory \cite{Lorentz1, Lorentz2, Lorentz3, Lorentz4, Lorentz5, Lorentz6, Lorentz7, bumblebee, bumblebee1}. Such scenarios arise naturally in various extensions of GR and high-energy theories, including string theory and modified gravity models.

An implementation of the latter mechanism arises from a non-minimal coupling with the Ricci scalar, where the second-order Kalb-Ramond (KR) tensor field \cite{KRoriginal} spontaneously breaks Lorentz symmetry, acquiring a nonzero vacuum expectation value \cite{VLorentz,VLorentz3}. This framework has been explored in various contexts \cite{KRteste,KRinfla, KRlenteforte, KRKumar, KRparticulas, KRparity, KRdarkmatter, KReletrico, KRtermo, PetrovX, PetrovX2, PetrovX3}.
In this work, we focus on analyzing the exact static and spherically symmetric solution for electrically charged black holes, first obtained in \cite{KReletrico}, within a self-interacting KR field coupled in a non-minimal way to the Einstein-Hilbert action \cite{VLorentz, KRMaluf}, together with an electromagnetic field.

Our goal is to further analyze the feasibility of KR gravity by studying the free-fall motion of a test particle in the electrically charged metric and the tidal forces it experiences. In Schwarzschild spacetime, a test particle falling from infinity reaches the speed of light at the event horizon, as observed by an external observer \cite{Crawford:2001jh}. However, for a static observer inside the horizon, its velocity can decrease to a minimum after crossing it, reflecting the black hole interior's expansion in the $t$-direction, resembling a cosmological scenario \cite{queda1, cenariocosmologico}. In the Reissner-Nordström (RN) and Kerr space-times, an additional contraction phase due to inner and outer horizons causes frequency shifts in emitted signals, redshift during expansion and blueshift during contraction, resembling a Doppler effect. The Doppler shift of electromagnetic signals exchanged between static and moving observers has been studied for Schwarzschild spacetime \cite{Doppler1} and extended to Anti-de Sitter and RN black holes \cite{Doppler2}. More broadly, the behavior of electromagnetic signals near horizons and redshift characteristics were examined in \cite{Doppler3}.

One of the most distinctive gravitational effects near black holes are tidal forces \cite{Inverno}, which arise from variations in the gravitational force and have been linked to supermassive black hole formation \cite{marestar, marestar2}. While well understood in the Schwarzschild space-time \cite{Inverno, chandra, Thorne}, they exhibit notable differences in charged spacetimes. In RN black holes, tidal forces can change sign or vanish at the event horizon, as first shown in \cite{Crispino}. Similar studies on electrically charged Hayward regular black holes \cite{CrispinoH} and those from nonlinear electrodynamics \cite{NEDtidal} reveal a dependence on the charge-to-mass ratio. Black bounces feature tidal peaks outside the horizon and sign changes within it, influenced by dust and radiation parameters \cite{BBtidal}. Tidal effects in effective matter fields, or ``dirty'' black holes, were also examined in \cite{Caio1}. Building on these studies \cite{Keertidal, holographictidal, 4Dtidal, SHtidal}, we investigate tidal effects on a freely falling body in a spacetime with spontaneous Lorentz symmetry-breaking (charged KR black hole).

This work is organized as follows. Section \ref{sec1} presents a concise derivation and analysis of the radial geodesic equation, which will be used throughout the paper. In Sec. \ref{sec2}, we examine the motion of a test particle in free fall toward the center of the charged KR black hole, focusing on how the parameter $l$ influences its velocity near the event horizon as observed from different reference frames.
In Sec.\;\ref{sectidalforce}, we derive the tidal force equations for a spherically symmetric, static metric and analyze how the charged KR space-time influences a neutral test object, revealing its dependence on the Lorentz symmetry-breaking parameter and black hole charge.
In Sec.\;\ref{sec5}, we solve the geodesic equation under two specific conditions and analyze the test particle's geodesic deviation, comparing the effects of $l$ and $q$ with those in the RN case. In Sec.\;\ref{sec6} we apply this method to two other space-times with Lorentz symmetry-breaking and compare the results. Finally, in Sec.\;\ref{Sec:Conclusion}, we summarize and conclude our work. Geometrized units ($G=1,c=1$) and the metric signature ($-,+,+,+$) are assumed.

\section{Electrically charged solution for the KR field}\label{sec00}

In this approach, we present a self-interacting KR field in the non-minimally coupled Einstein-Hilbert action \cite{KR, KRMaluf}  and with Maxwell field interaction \cite{KReletrico}.   The action of the theory is
\begin{eqnarray}
&& S=\int d^4x\sqrt{-g}\Bigg[ R-\frac{1}{6}H^{\mu\nu\rho}H_{\mu\nu\rho}-V(B^{\mu\nu}B_{\mu\nu})\nonumber\\
&& +\zeta_2 B^{\rho\mu}B^{\nu}_{\;\;\mu}R_{\rho\nu}+\zeta_3 B^{\mu\nu}B_{\mu\nu}R - \frac{1}{2}F^{\mu\nu}F_{\mu\nu}\nonumber\\
&&-\eta B^{\mu\nu}B^{\alpha\beta} F_{\mu\nu}F_{\alpha\beta}\Bigg]\,.    \label{açao1}
\end{eqnarray}
Here $H_{\mu\nu\rho}=\partial_{[\mu}B_{\nu\rho]}$ where $B_{\mu\nu}$ is the KR field strength tensor  and $F_{\mu\nu}=\partial_\mu A_\nu-\partial_\nu A_\mu$  the electromagnetic field strength one. If we consider the expectation value of the field contraction in vacuum as $\langle B_{\mu\nu}\rangle =b_{\mu\nu}, \langle B^{\mu\nu}B_{\mu\nu}\rangle =\mp b^2$, the coupling term $\zeta_3$ in the action is reduced to a linear term in $R$, which can be converted into an Einstein-Hilbert term by a coordinate transformation. Furthermore, the KR field can be decomposed into $B_{\mu\nu}=\hat{E}_{[\mu}v_{\nu]}+\epsilon_{\mu\nu\alpha\beta}v^{\alpha}\hat{B}^{\beta}$, where $v^{\alpha}$ is a time-like quadri-vector, and $\hat{E}^{\mu}$ and $\hat{B}^{\mu}$ are the space-like electric and magnetic pseudo-fields.   
For this solution, we consider an ansatz of the potential vector $A_\mu=-\Phi(r)\delta^t_\mu$. For consistency of the solution, it is necessary to include the last interaction term in \eqref{açao1} through the coupling constant $\eta$. When the KR field acquires an expected value in vacuum other than zero, the Lagrangian triggers the breaking of Lorentz symmetry in the electromagnetic sector, enabling the existence of electrically charged solutions. \cite{KReletrico}. 
 If we now vary the action with respect to the metric, we obtain
\begin{eqnarray}
&& R_{\mu\nu}-\frac{1}{2}g_{\mu\nu}R=\frac{1}{2}H_{\mu\alpha\beta}H_{\nu}^{\;\;\alpha\beta}-\frac{1}{12}g_{\mu\nu}H^{\alpha\beta\rho}H_{\alpha\beta\rho}\nonumber\\
&& +2V'(X)B_{\alpha\mu}B^{\alpha}_{\;\;\nu}-g_{\mu\nu}V(X)+\zeta_2\Bigg[\frac{1}{2}g_{\mu\nu}B^{\alpha\rho}B^{\beta}_{\;\;\rho}R_{\alpha\beta}\nonumber\\
&& -B^{\alpha}_{\;\;\mu}B^{\beta}_{\;\;\nu}R_{\alpha\beta}-B^{\alpha\beta}B_{\nu\beta}R_{\mu\alpha}-B^{\alpha\beta}B_{\mu\beta}R_{\nu\alpha}\nonumber\\
&&  +\frac{1}{2}\nabla_{\alpha}\nabla_{\mu}\left(B^{\alpha\beta}B_{\nu\beta}\right)+\frac{1}{2}\nabla_{\alpha}\nabla_{\nu}\left(B^{\alpha\beta}B_{\mu\beta}\right)\nonumber\\
&& -\frac{1}{2}\nabla^{\alpha}\nabla_{\alpha}\left(B_{\mu}^{\;\;\rho}B_{\nu\rho}\right)-\frac{1}{2}g_{\mu\nu}\nabla_{\alpha}\nabla_{\beta}\left(B^{\alpha\rho}B^{\beta}_{\;\;\rho}\right)\Bigg ]  \nonumber\\
&&+2F_{\mu\alpha}F^{\;\;\alpha}_{\nu}-\frac{1}{2}g_{\mu\nu}F^{\alpha\beta}F_{\alpha\beta}\nonumber\\
&&+\eta\left(8B^{\alpha\beta}B^{\;\;\sigma}_{\nu}F_{\alpha\beta}F_{\mu\sigma}-g_{\mu\nu}B^{\alpha\beta}B^{\sigma\rho}F_{\alpha\beta}F_{\sigma\rho}\right)\,.\label{movKR1}   
\end{eqnarray}
The variation of action \eqref{açao1} with respect to the KR field gives us the following equation of motion:
\begin{eqnarray}
\nabla^{\alpha}H_{\alpha\mu\nu}+&&3\zeta_2 R_{\alpha[\mu}B^{\alpha}_{\;\;\nu]}\nonumber\\
&&-6V'(X)B_{\mu\nu}-12\eta B^{\alpha\beta}F_{\alpha\beta}F_{\mu\nu}=0\,.\label{movKR2}
\end{eqnarray}
The functional variation with respect to $A^{\mu}$ gives the modified Maxwell equations as
\begin{eqnarray}
\nabla^{\mu}\left(F_{\mu\nu}+2\eta B_{\mu\nu}B^{\alpha\beta}F_{\alpha\beta}\right)=0\,,\label{movKR3}
\end{eqnarray}
so that when $\eta\rightarrow 0$ we recover the standard Maxwell equations. We shall assume spherical and static symmetry, so the metric is given by 
\begin{eqnarray}
ds^2=-A(r)dt^2+B(r)dr^2+r^2\left(d\theta^2+\sin^2\theta d\phi^2\right)\,.\label{metricageral}    
\end{eqnarray}
For a solution with only $\hat{E}\equiv \hat{E}(r)=b_{10}$, we have the 2-form $b_{(2)}=-\hat{E}(r)dt\wedge dr$. In this particular case, we have $H_{\mu\nu\alpha}\equiv 0$, or $H_{(3)}=db_{(2)}=0$. Taking $V(x)=\lambda X^2/2$, where $X=B^{\mu\nu}B_{\mu\nu}+b^2$, we have $V'\equiv 0$. Therefore $\hat{E}(r)=|b|\sqrt{A(r)B(r)/2}$ and $b^{\mu\nu}b_{\mu\nu}=-b^2$, with that, the equations of motion become
\begin{eqnarray}
&& 2\frac{A''}{A} -\frac{A'B'}{AB}-\frac{A'^2}{A^2} +\frac{4A'}{rA}-\frac{4\left(1-2\eta b^2\right)\Phi'^2}{(1-l)A}=0\,,\;\;\;\label{eqmo1}\\
&& 2\frac{A''}{A} -\frac{A'B'}{AB}-\frac{A'^2}{A^2} -\frac{4B'}{rB}-\frac{4\left(1-2\eta b^2\right)\Phi'^2}{(1-l)A}=0\,,\;\;\;\label{eqmo2}\\
&&2\frac{A''}{A} -\frac{A'B'}{AB}-\frac{A'^2}{A^2} +\frac{(1+l)}{lr}\left(\frac{A'}{A}-\frac{B'}{B}\right)\nonumber\\
&& -2\frac{B}{lr^2}+\frac{2(1-l)}{lr^2}-2\frac{\left(1-6\eta b^2\right)\Phi'^2}{lA}=0=0\label{eqmo3}\,,
\end{eqnarray}
where we define $l=\zeta_2 b^2/2$. From Eqs.\eqref{movKR2} and  \eqref{movKR3} we obtain the following two equations 
\begin{eqnarray}
2\frac{A''}{A}-\frac{A'^2}{A^2}+\frac{2}{r}\left(\frac{A'}{A}-\frac{B''}{B}\right)-\frac{A'}{A}\frac{B''}{B}-\frac{8\eta b^2\Phi'^2}{lA}=0\,,\;\;\;\;\;\;\;\label{movKRMax1}\\
\left(1-2\eta b^2\right)\left[\Phi''+\frac{\Phi'}{2}\left(\frac{4}{r}-\frac{A'}{A}-\frac{B'}{B}\right)\right]=0\,.\;\;\;\;\;\;\;\label{movKRMax2}
\end{eqnarray}
respectively.  Subtracting the first equation from the second one we get
\begin{eqnarray}
\frac{A'}{A}+\frac{B'}{B}=0\,,    
\end{eqnarray}
whose solution is
\begin{eqnarray}
 B(r)=\frac{c_1}{A(r)}\,.   \label{Br}
\end{eqnarray}
Substituting this solution into \eqref{movKRMax2} we obtain (taking $c_1\rightarrow 1$), 
\begin{eqnarray}
\Phi''+\frac{2}{r}\Phi'=0
\end{eqnarray}
where we obtain the electrostatic potential
\begin{eqnarray}
\Phi(r)=\frac{c_2}{r}+c_3\,,
\end{eqnarray}
taking $\Phi(\infty)=0$, then $c_3=0$.   In this case, the conserved current takes the modified form
\begin{eqnarray}
J^\mu=\nabla\left(F^{\mu\nu}+2\eta B^{\mu\nu}B^{\alpha\beta}F_{\alpha\beta}\right)\,,
\end{eqnarray}
Therefore, the electric charge determined via Stokes's theorem is
\begin{eqnarray}
q&=&-\frac{1}{4\pi}\int_{\partial\Sigma}d\theta d\phi\sqrt{\gamma^{(2)}}n_\mu\sigma_\nu\left(F^{\mu\nu} +2\eta B^{\mu\nu}B^{\alpha\beta}F_{\alpha\beta}\right)\nonumber\\
&=&\left(1-2b^2\eta\right)c_1\,,
\end{eqnarray}
where we take the normal vectors $n_\mu=(1,0,0,0)$ associated with a 3-dimensional spacelike region $\Sigma$;  $\sigma_\mu=(0,1,0,0)$ associated with $\partial\Sigma$ and $\gamma^{(2)}=r^2(d\theta^2+\sin^2\theta d\phi^2)$, to obtain the integration constant $c_2$, which gives us the electrostatic potential 
\begin{eqnarray}
\Phi(r)=\frac{q}{(1-2b^2\eta)r}\,.\label{Phifinal}
\end{eqnarray}
Using \eqref{Br} and \eqref{Phifinal} and substituting into \eqref{eqmo3}, and integrating we have
\begin{eqnarray}
    A(r)=\frac{c_4}{1-l}+\frac{c_5}{r}+\frac{q^2}{(1-l)^2r^2}\,.
\end{eqnarray}
where we take $\eta=l/2b^2$ for the consistency of the solution. The constant $c_5$ can be calculated from the mass of Komar, which leads to $c_5=-2M$. To revert to the Schwarzschild solution, if we have the coupling with the zero KR field, $l=0$ and $q=0$, we must then have $c_4=1$, so that the line element takes the form:
\begin{eqnarray}
ds^2 =  -A(r)dt^2+A^{-1}(r)dr^2 +r^2 (d\theta^2+\sin^2\theta d\phi^2)\,,\;\;\;\;\;\;\; \label{SSSline}
\end{eqnarray}
with
\begin{eqnarray}
A(r)= \frac{1}{1-l}-\frac{2M}{r}+\frac{q^2}{(1-l)^2r^2}\,,\label{Ametric}
\label{KR0}
\end{eqnarray}
where $l$ is an additional parameter that characterizes the spontaneous Lorentz symmetry-breaking  and $q$ is the electric charge. Note that, in the limit $r\rightarrow\infty$, the metric \eqref{Ametric} is not Minkowskian. Note also that the solution only admits values $-1<l<1$ in order to maintain the original signature of the space-time\footnote{Note that a transformation in the radial coordinate from $\tilde{r}\rightarrow r\sqrt{1-l}$ results in a metric that can be viewed as the RN metric with a solid angle deficit.}.

 The above line element resembles the causal structure of the charged RN solution of GR, with horizons located at
\begin{eqnarray}
r_{\pm}=(1-l)\left[M \pm \sqrt{M^2-\frac{q^2}{(1-l)^3}}\right]\,,
\end{eqnarray}
with $r_+$ the event horizon and $r_-$ the Cauchy one. 
Furthermore, the analysis of \cite{KReletrico} constrains the parameter $l$ using the inference of the shadow's radius of Sgr A* by the EHT Collaboration to 
\begin{equation}
-4.59\times 10^{-3} \leq l \leq 1.24\times 10^{-1}\,.
\end{equation}

In this context, several of the present authors recently investigated gravitational lensing  \cite{nosso2} in a Schwarzschild-type black hole spacetime within the KR background field  proposed in \cite{KR}, analyzing the deflection angle and key observables such as image position, luminosity, and delay time. Additionally, in \cite{nosso3}, we examined time-like geodesics in this background, revealing significant modifications to orbital momentum and energy depending on the parameter $l$, with applications to extreme mass ratio inspiral (EMRI) systems and waveform characteristics of periodic orbits. Furthermore, we studied time-like and light-like geodesics in KR gravity around a black hole to constrain the Lorentz symmetry-breaking parameter, using the precession of the S2 star orbiting Sgr A* and geodesic precession around the Earth, ultimately providing a constraint on the spontaneous symmetry-breaking parameter in the  following interval \cite{nosso}:
\begin{equation}
    -0.185022 \leq l \leq 0.0609
\end{equation} 
in contrast with 
\begin{equation}
    -6.1\times 10^{-13} \leq l \leq 2.8\times 10^{-14}
\end{equation} 
obtained in \cite{KR} using the Shapiro time-delay effect. We shall use a larger positive value, $l = 0.1$, in order to illustrate more clearly, from a graphical perspective, the difference between the RN solution and the Kalb–Ramond electrically charged solution. However, the overall physical results are not altered by this choice.

\section{Radial Geodesic} \label{sec1}

The Lagrangian that describes a test particle in GR is  given by $\mathcal{L}=\frac{1}{2}g_{\mu\nu}\dot{x}^{\mu}\dot{x}^{\nu}$, where the overdot denotes differentiation with respect to the affine parameter. When a particle moves along time-like geodesics, it must satisfy the condition $\mathcal{L}=-1$, and therefore we have \cite{Inverno}
\begin{equation}
g_{\mu\nu}\dot{x}^{\mu}\dot{x}^{\nu}=-1.
\label{gmn}
\end{equation}
In a general static and spherically symmetric metric of the form (\ref{SSSline}) and considering radial geodesic motion, for which angular momentum is null and $\dot{\theta}= \dot{\phi}=0$ by assumption, then Eq. \eqref{gmn} reads
\begin{equation}
    -A(r)\dot{t}^2+A(r)^{-1}\dot{r}^2=-1\,.
    \label{eq:dsmov}
\end{equation}
In this scenario $E=A(r)\dot{t}$ is the energy of the particle  per unit mass and it is conserved. Replacing this energy conservation in Eq. \eqref{eq:dsmov} we find
\begin{eqnarray}
\dot{r}^2+A(r)=E^2\,. \label{rponto}
\end{eqnarray}

Let us assume a test particle that moves from rest in the radial position $r=b$. Then its rest energy is given simply by $E=\sqrt{A(r=b)}$.   The force that the black hole exerts on the test particle is calculated from the Newtonian radial acceleration as $\mathcal{A}^N\equiv\ddot{r}$, so that by using Eq. \eqref{rponto} we have 
\begin{eqnarray}
\mathcal{A}^N=-\frac{A'(r)}{2}\,,
\end{eqnarray}
where the prime denotes the derivative with respect to the radial coordinate. Therefore, the Newtonian radial acceleration provides the force that the charged KR black hole exerts on a neutral test particle in free fall, as given by 
\begin{eqnarray}
\mathcal{A}^N=\frac{1}{2} \left(\frac{2 q^2}{(1-l)^2 r^3}-\frac{2 M}{r^2}\right) .
\end{eqnarray}
In this expression, for  $l\rightarrow 0$ we recover the one of the RN black hole \cite{Crispino}, and additionally if $q\rightarrow 0$ we recover the usual Schwarzschild expression $\mathcal{A}^N=-M/r^2$. 

For a test particle falling freely from rest at $r=b$, there is an inflection point $R^{\rm stop}$ where the particle resumes its radial motion. This point is obtained from imposing the condition $E^2-A(R^{\rm stop})=0$, and it reads explicitly as
\begin{eqnarray}
R^{\rm stop}=\frac{b q^2}{2 b (l-1)^2 M-q^2}\,.
\end{eqnarray} 
In our case it is always located within the Cauchy horizon, similarly as in the RN case where $l\rightarrow 0$. Note that if the starting point of the test particle is located at asymptotic infinity, $b\rightarrow\infty$, then $R^{\rm stop}\rightarrow\frac{q^2}{2 (l-1)^2 M}$.  


\section{Free fall and Gravitational Doppler effect} \label{sec2}

The proper time of a test particle in free fall can be conveniently written as \cite{queda1, Doppler2} 
\begin{eqnarray}
d\tau^2 = -A(r)dt^2\left(1-v^2\right)\,,
\label{tau2}
\end{eqnarray}
where
\begin{eqnarray}
v^2=\frac{1}{A(r)^2}\left(\frac{dr}{dt}\right)^2 \,, \label{v1}
\end{eqnarray}
is the velocity of the particle relative to a static observer. Since $E$ is the single constant of free-fall radial geodesic motion, using Eq.\,\eqref{v1} the only non-zero component of the particle's velocity measured by a static observer located at $r$ is given by
\begin{eqnarray}
v^2=\frac{E^2-A(r)}{E^2}\,,
\label{Vext}
\end{eqnarray}
and which tends to the speed of light at the horizon. Note that, while for asymptotically-flat (Minkowiski) space-times, $A(r)\rightarrow 1$ when $r\rightarrow \infty$ and thus $E\rightarrow 1$ in the charged KR space-time described by Eq.\,\eqref{KR0}, $A(r\rightarrow\infty)\rightarrow 1/(1-l)$ so $E^2\rightarrow 1/(1-l)$. 

Consider the two horizons, $r_+$ and $r_-$, of the KR black hole. When crossing the outer horizon $r$ plays the role of the time coordinate $t$, since the time-like Killing vector is converted into a space-like one. Therefore, for an observer located between the two horizons, i.e. $r_-<r_0<r_+$, the velocity of the test particle in the free-fall frame between the horizons is measured as
\begin{eqnarray}
V^2=\frac{E^2}{E^2-A(r)}\,.\label{Vint}
\end{eqnarray}
Therefore, from Eqs. \eqref{Vext} and \eqref{Vint} in the charged KR space-time, the velocities for $b>r_+$ and $r_-< r_0 <r_+$ are,
\begin{eqnarray}
v^2=\frac{2 (1-l)^2 M r-q^2}{(1-l) r^2}\,,\label{Vext2}\\ V^2=\frac{(1-l) r^2}{2 (1-l)^2 M r-q^2}\,, \label{Vint2}
\end{eqnarray}
respectively. 

\begin{figure}[ht!]
\centering
{\includegraphics[width=8.5cm]{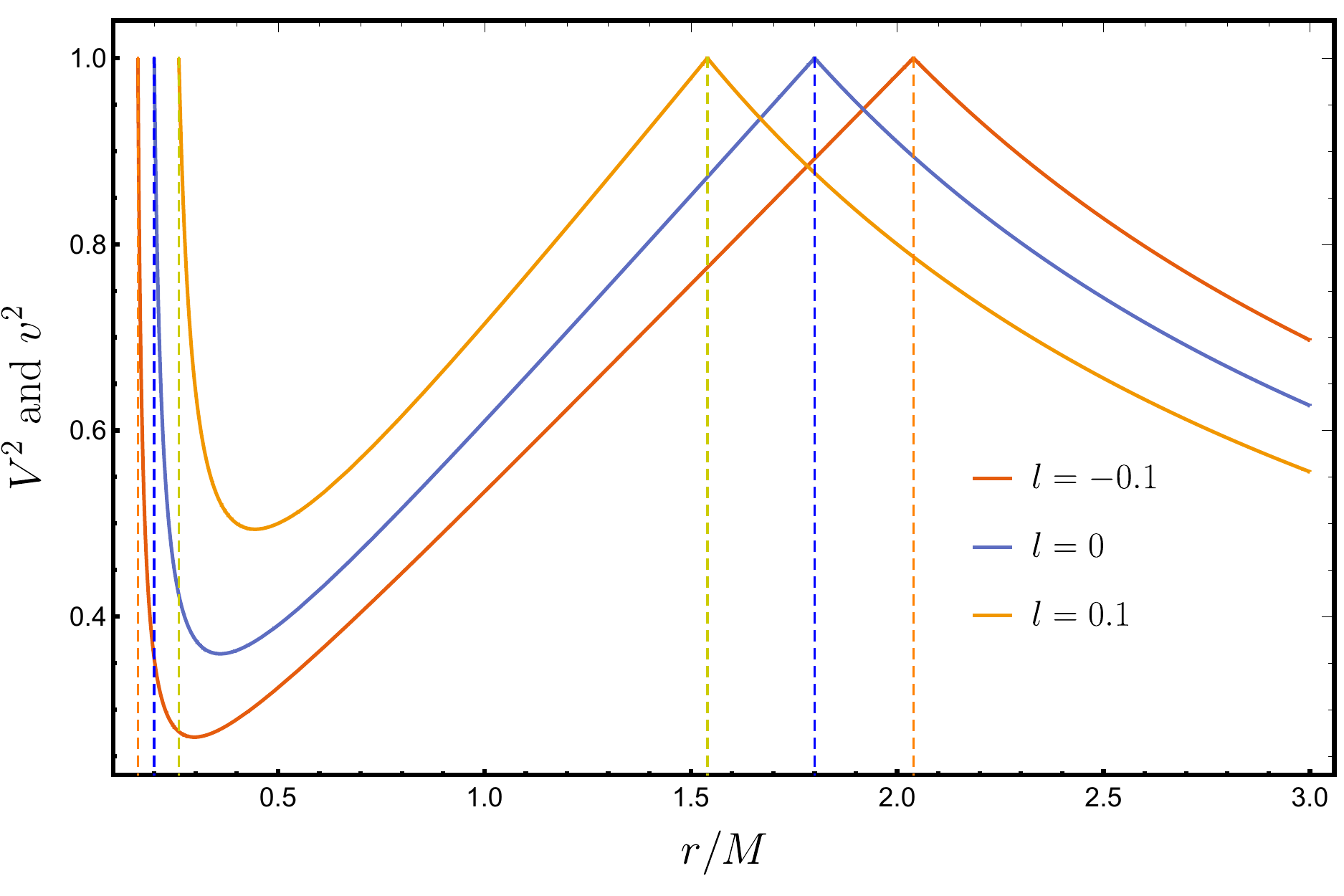}}
\caption{Velocities reached by the test particle in free fall, outside the outer horizon $(v)$ towards the center of the black hole and between the horizons $(V)$. We took $q=0.6$.}\label{Fig1}
\end{figure}

In Fig.\,\ref{Fig1},  we plot both velocities \eqref{Vext2} and \eqref{Vint2}. We see that  when $r\rightarrow r_{\pm}$ both approach the speed of light in vacuum. 
The dashed lines mark the location of the horizons for each value of the spontaneous Lorentz symmetry-breaking parameter, separating the inside and outside regions. When approaching the outer horizon from infinity, the test particle has a velocity measured by the observer as $b>r_+$, approaching the speed of light in vacuum. When crossing the outside horizon, an internal observer  between $r_+$ and the Cauchy horizon $r_-$ registers a decrease in the velocity of the particle up to a minimum located at $r_{\rm min}=\frac{q^2}{(1-l)^2 M}$ followed by an increase to the speed of light at $r_-$. Note that the minimum of $V$ is modified due to the initial energy of the particle being altered by $l$. 

Let us consider now an observer in the free-fall frame of reference near the charged KR black hole which exchanges electromagnetic signals with a second external observer at $r=b$. The tangent four-vector of the photon in this case is $\kappa^\mu=dx^\mu/d\sigma$ and satisfies $g_{\mu\nu}\kappa^\mu\kappa^\nu=0$ where $\sigma$ is the affine parameter of the photon world line. From the conservation of energy, the components of the wave vector are therefore, $\kappa^0=\omega_\infty/A(r)$ and $\kappa^1=\mp\omega_\infty$, with $\mp$ representing photon entry and exit. Thus, the frequency ratio between the signals sent, $\omega^s_{rf}$  by the observer of the reference in free fall and received, $\omega^r_b$, by the fixed observer at $b$ is given by 
\begin{eqnarray}
\frac{\omega^r_b}{\omega^s_{rf}}=\frac{(\kappa^\mu u_\mu)_r}{(\kappa^\mu u_\mu)_s}=1-\sqrt{\frac{E^2-A(r)}{E^2}} = 1-v \,,\label{r/e}
\end{eqnarray}
where we have used Eqs.\,\eqref{rponto} and \eqref{tau2}. As $v\rightarrow 1$ at the horizon, this means via Eq.\,\eqref{r/e} that the photon received by the outside observer is infinitely redshifted. On the other hand, the frequency ratio of the signals emitted by the observer in $b$ and received by the observer in the free-fall reference frame is 
\begin{eqnarray}
\frac{\omega^r_{rf}}{\omega^s_b}=\frac{1}{1+v}\,.\label{r/e2}
\end{eqnarray}
Note that when $v\rightarrow 1$ (free-falling particle approaches the outside horizon) the ratio \eqref{r/e2} goes to $ 1/2$, so clearly there is an asymmetry in the sending of signals. This reflects the fact that the observer in $b$ receives signals emitted by the observer in the free-fall frame only until it crosses the horizon $r_+$, while the free-falling observer continuously receives signals emitted by the observer in $b$ even after it crosses the horizon.  From the perspective of an observer outside the horizon $r>r_+$
 the gravitational redshift inside the outside horizon becomes infinite, because the signals emitted inside $r_+$ can never escape.

Inside the horizon, the coordinates $t$ and $r$ reverse their roles due to the change in the metric signature, which implies that the motion towards the inner horizon unavoidably both for particles and for light.  The signals recorded by a given observer at $r_0$, located between the inside and outside horizons, provide the relationship between the frequencies in this region as \cite{queda1}
\begin{eqnarray}
\frac{\omega^r_{r_{0}}}{\omega^s_{rf}}=\left(1+V^{-1}\right)^{-1}=\frac{1}{\sqrt{\frac{2 (1-l)^2 M r-q^2}{(1-l) r^2}}+1} \,.
\end{eqnarray}

\begin{figure}[t!]
\centering
{\includegraphics[width=8.5cm]{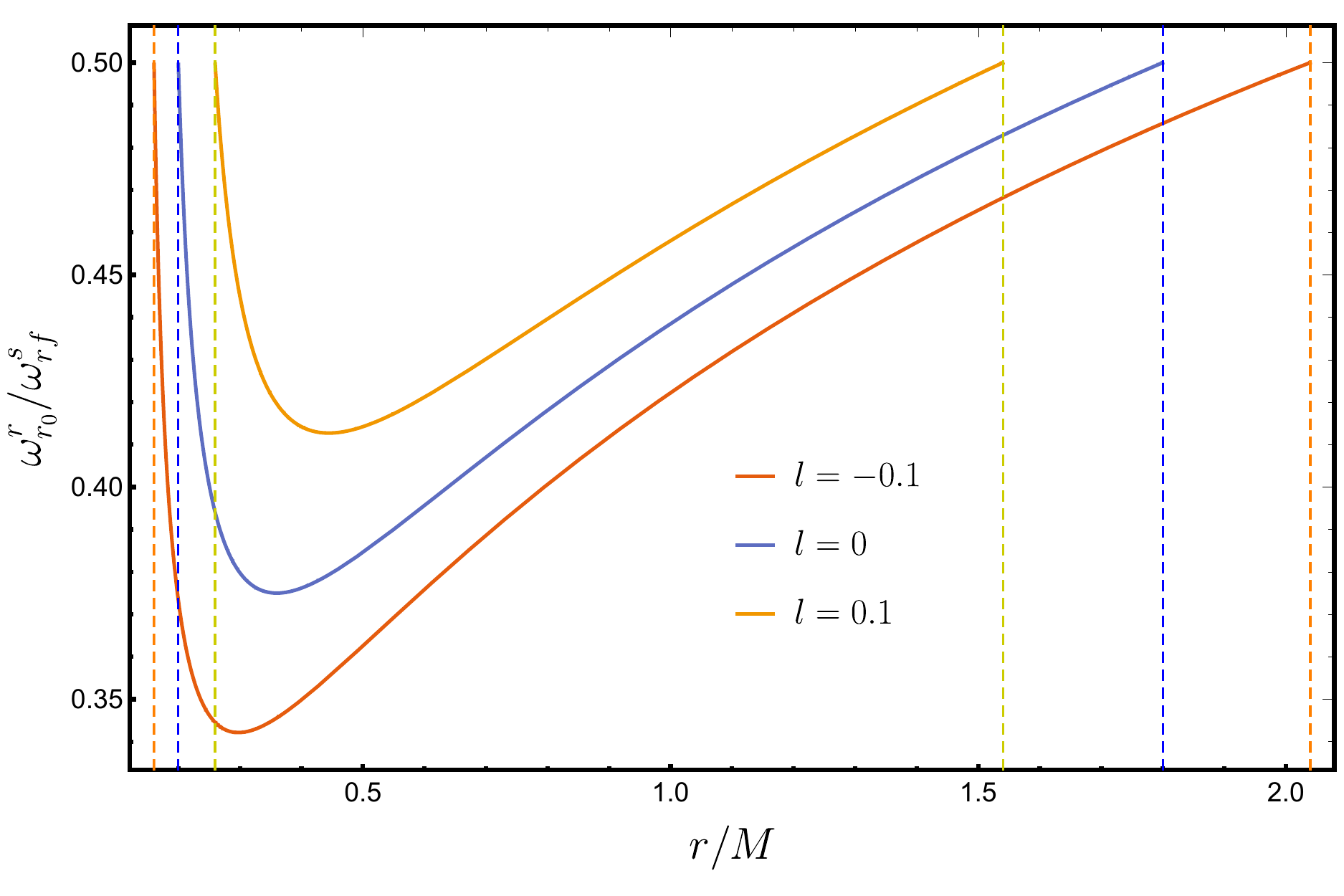} }
\caption{Ratio between the frequencies of electromagnetic signals emitted by a free-falling test particle and received by a fixed observer between the $r_{\pm}$ horizons. Again $q=0.6$. }\label{Fig2}
\end{figure}

This ratio is plotted in Fig.\,\ref{Fig2} for different values of $l$. As the free-falling particle passes through the horizon $r_+$ towards the inner region, the space-time is expanding and the observer will perceive a Doppler redshift effect to a minimum at $r_m=\frac{q^2}{(1-l)^2 M}$. From $r_m$ the spacetime contracts to the inner horizon $r_-$ where a Doppler blueshift effect takes place \cite{Doppler1, Doppler2}. The ratio between the frequencies sent and received tends to  $1/2$ when $r\rightarrow r_{\pm}$, which implies $V\rightarrow c$. Here, as with the velocity between horizons, $l$ modifies the value of $r_m$ due to the change in initial energy.  

\section{Tidal Forces}\label{sectidalforce}

\subsection{Tidal forces for static and spherically symmetric spacetimes}

The tidal forces experienced by a free-falling particle due to the black hole's gravitational field is defined from the spatial components of the geodesic deviation vector $\eta^\mu$. This vector describes the infinitesimal distance between two nearby particles in free fall and can be obtained from the geodesic deviation equation \cite{Inverno}
\begin{eqnarray}
\frac{D^2\eta^\mu}{D\tau^2}+R^{\mu}_{\,\,\nu\alpha\beta}v^\nu\eta^\alpha v^\beta = 0\,,
\end{eqnarray}
where $R^{\mu}_{\,\,\nu\alpha\beta}$ is the Riemann tensor and $v^\beta$ is the unit vector tangent to the geodesic.  For a test body moving along a geodesic path in space-time, each point follows its own distinct geodesic. This results in variations in acceleration between neighboring geodesics, primarily causing the body to experience an elongation and compression due to the tidal force. This force is described from a set of bases (called tetrads) of local coordinates in the frame of the object that forms a local frame of reference at each point of the geodesic. The components of such tetrads $e^\mu_{\hat{a}}=(e^\mu_{\hat{0}}, e^\mu_{\hat{1}}, e^\mu_{\hat{2}}, e^\mu_{\hat{3}})$ are found in our case as
\begin{eqnarray}
e^\mu_{\hat{0}} &=& \left(\frac{E}{A(r)}, -\sqrt{E^2-A(r)}, 0, 0\right)\;,\nonumber\\ e^\mu_{\hat{1}}&=&\left(-\frac{\sqrt{E^2-A(r)}}{A(r)}, E, 0, 0\right)\,,\nonumber\\
e^\mu_{\hat{2}}&=&\left(0, 0, \frac{1}{r}, 0 \right)\,,\,\,\, e^\mu_{\hat{3}}=\left(0, 0, 0, \frac{1}{r\sin\theta}\right)\,,
\end{eqnarray}
satisfying the orthonormality condition $e^\mu_{\hat{a}}e_{\mu\hat{b}}=\eta_{\hat{a}\hat{b}}$ where $ \eta_{\hat{a}\hat{b}}={\rm diag}\left(-1, 1, 1, 1\right)$ is the Minkowski metric and $e^\mu_{\hat{0}}=v^\mu$ is the four-velocity. The geodesic deviation vector can be expanded into $\eta^\mu=e^\mu_{\hat{a}}\eta^{\hat{a}}$ and fix the time component as $\eta^{\hat{0}}=0$.

The Riemann tensor can be calculated on a tetrad basis via
\begin{eqnarray}
R^{\hat{a}}_{\,\,\hat{b}\hat{c}\hat{d}}=e^{\hat{a}}_{\,\,\mu}e^\nu_{\,\,\hat{b}}e^\rho_{\,\,\hat{c}}e^\sigma_{\,\,\hat{d}}R^\mu_{\,\,\nu\rho\sigma}\,,
\end{eqnarray} 
whose nonzero components are
\begin{eqnarray}
&& R^{\hat{0}}_{\,\,\hat{1}\hat{0}\hat{1}}=-\frac{A''(r)}{2}\,,\label{R0101}\\
&& R^{\hat{0}}_{\,\,\hat{2}\hat{0}\hat{2}}=R^{\hat{0}}_{\,\,\hat{3}\hat{0}\hat{3}}=R^{\hat{1}}_{\,\,\hat{2}\hat{1}\hat{2}}=R^{\hat{1}}_{\,\,\hat{3}\hat{1}\hat{3}}=-\frac{A'(r)}{2r}\,,\label{R0202}\\
&&R^{\hat{2}}_{\,\,\hat{3}\hat{2}\hat{3}}=\frac{1-A(r)}{r^2}\,.\label{R2323}
\end{eqnarray}
Since $e^\mu_{\,\,\hat{a}}$ are transported in parallel along the geodesic, from Eq.\,\eqref{R0101}--\eqref{R2323} the tidal force equations for a neutral object in free fall near the black hole are given by 
\begin{eqnarray}
&&\frac{D^2\eta^{\hat{1}}}{D\tau ^2}=-\frac{A''(r)}{2}\eta^{\hat{1}}\,,\label{radial1}\\
&&\frac{D^2\eta^{\hat{i}}}{D\tau ^2}=-\frac{A'(r)}{2r}\eta^{\hat{i}}\,,\label{angulari}
\end{eqnarray}
where $\hat{1}$ is the radial component and $i=2, 3$ are the angular components $(\hat{\theta},\hat{\phi})$ of the tidal force in the local frame. 

In the following sections we will focus on the study of tidal forces exerted on a neutral object in free fall near the black hole described by the metric \eqref{KR0} of the charged KR black hole.

\subsection{Radial tidal forces in KR black hole}

\begin{figure*}[ht!]
\centering
\subfigure[$q = 0.3$] 
{\label{Fradiala}\includegraphics[width=8.5cm]{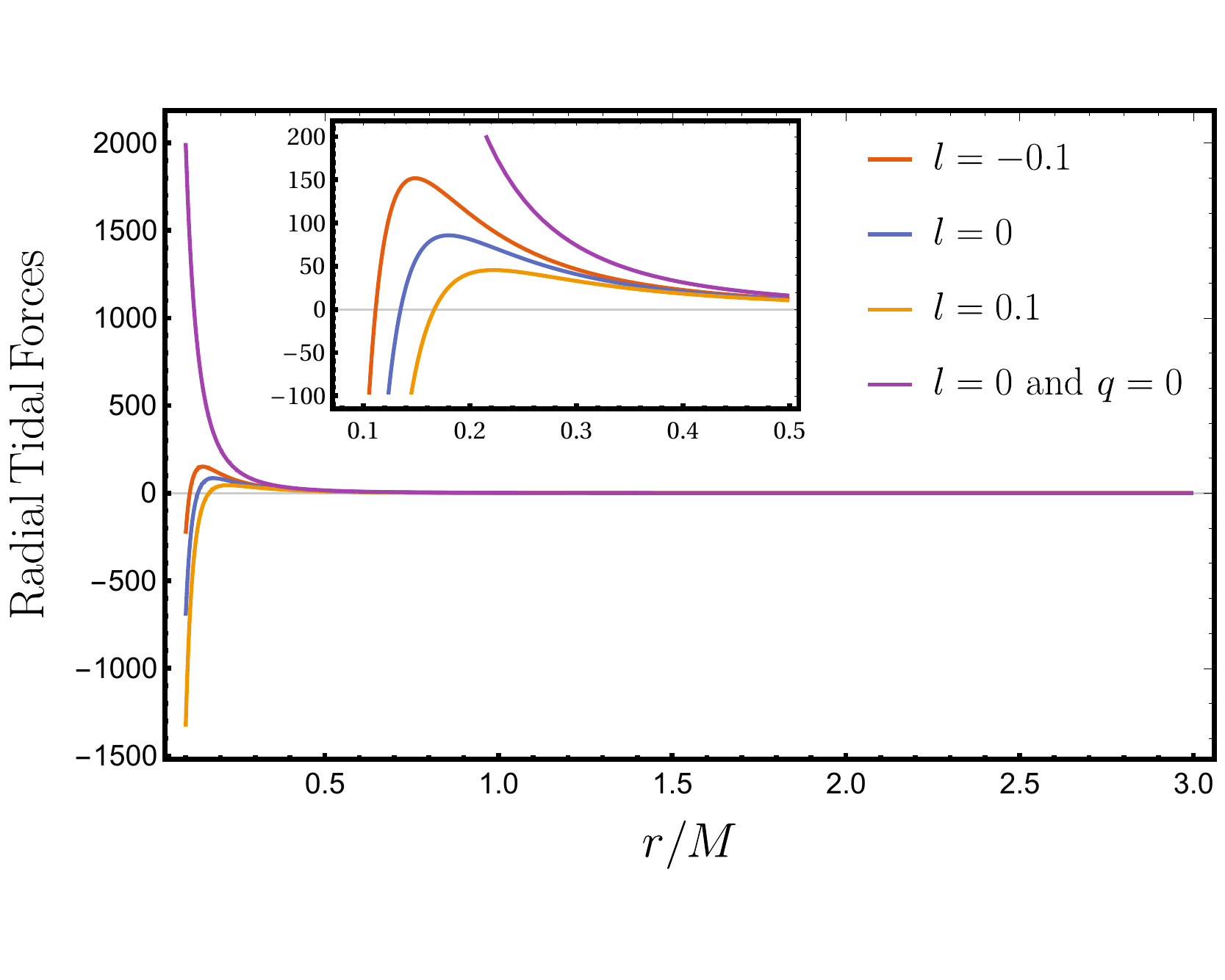} }
\hspace{0.1cm}
\subfigure[ $l=1.24\times 10^{-1}$] 
{\label{Fradialb}\includegraphics[width=8.5cm]{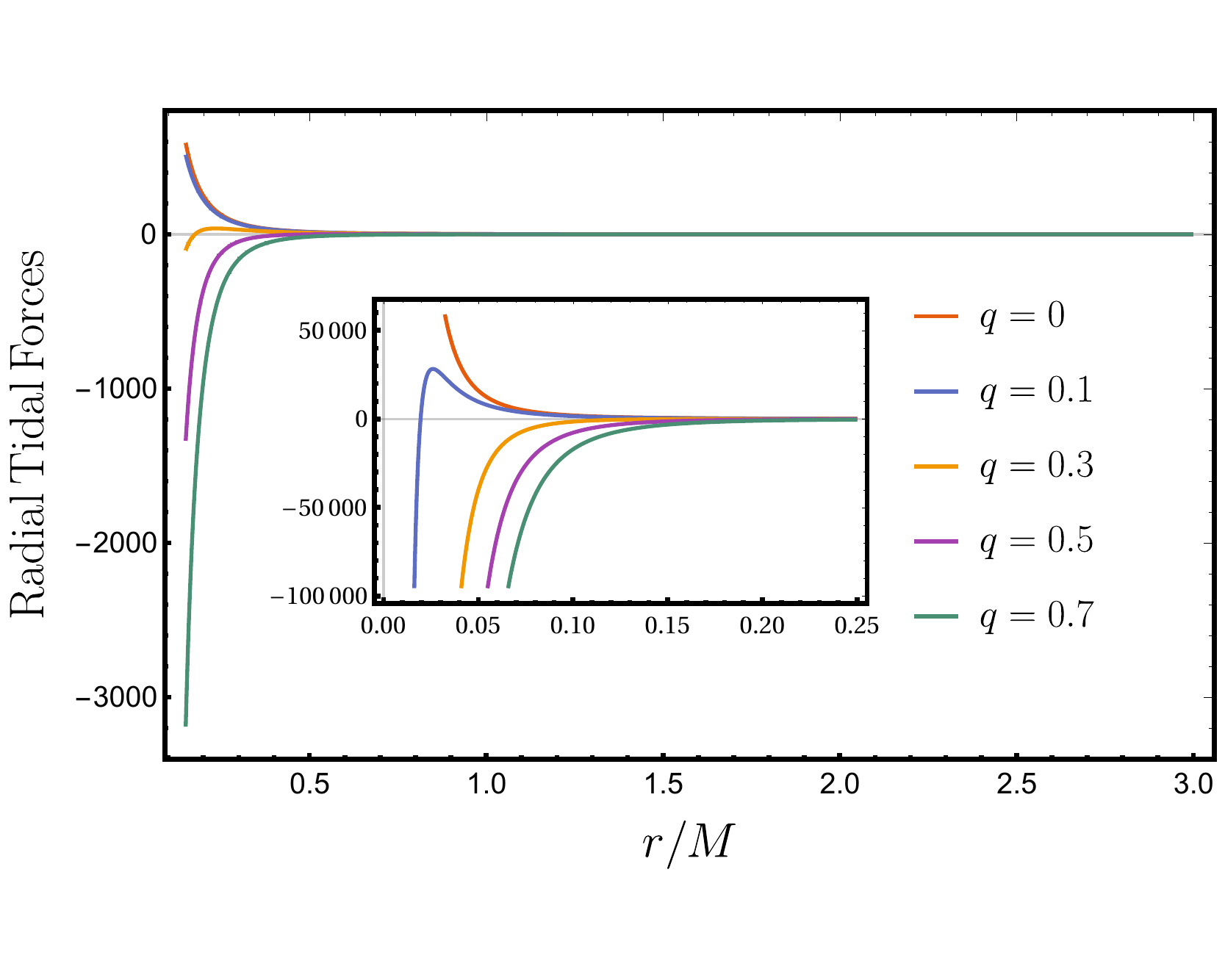} }
\caption{(a) Radial tidal forces with $q=0.3$ and different $l$.  (b) Radial tidal forces with $l=1.24\times 10^{-1}$ for different $q$. Here we have not used the negative values of $l$ because within the range obtained in \cite{KReletrico} its effect is very small and its effects are not evident. }\label{Fig3}
\end{figure*}  

Using Eq.\,\eqref{radial1} applied to the metric \eqref{KR0} we obtain the radial tidal force in the charged KR black hole  as
\begin{eqnarray}
\frac{D^2\eta^{\hat{1}}}{D\tau ^2}=\frac{1}{2} \left(\frac{4 M}{r^3}-\frac{6 q^2}{(1-l)^2 r^4}\right)\eta^{\hat{1}}\,,\label{Fradial}
\end{eqnarray}
which recovers the result for RN when $l\rightarrow 0$ \cite{Crispino} and the one of Schwarzschild when $q=0$. 
Unlike in the Schwarzschild solution 
 \cite{Inverno, holographictidal, SHtidal}, the radial tidal force can be zero for a given $r=R^{\rm rad}_0$, which is obtained from Eq. \eqref{Fradial} as
\begin{eqnarray}
R^{\rm rad}_0=\frac{3 q^2}{2 (1-l)^2 M}\,,\label{radial0}
\end{eqnarray}
while it takes its maximum at
\begin{eqnarray}
R^{\rm rad}_{\rm max}=\frac{2 q^2}{(1-l)^2 M}\,.\label{radialmax}
\end{eqnarray}

  To understand how spontaneous symmetry-breaking affects the radial tidal force,  in Fig.\,\ref{Fradiala} we plot Eq.\,\eqref{Fradial} for the charged KR black hole with different $l$ and fixed $q=0.3$. Similarly to the results found in the RN space-time \cite{Crispino}, there is always a maximum for any $q\neq 0$ in $R^{\rm rad}_{\rm max}$, given by Eq.\,\eqref{radialmax}. However, $R^{\rm rad}_{\rm max}$ is shifted (left) near $r=0$ when $l<0$ and (right) away from $r=0$ when $l>0$. There is therefore a reversal in the radial elongation force for compression when $l\neq 0$. The figure also shows that the radial tidal force is stronger when $l<0$ and lower when $l>0$.   Furthermore for $q=0$ and $l=0$ (the Schwarzschild case),  the radial force is positive and diverges as the object falls rapidly from infinity and approaches $r=0$, having only an elongation behaviour, an effect known as \textit{spaghettification}.
 In Fig.\,\ref{Fradialb} we plot the radial force for $l=1.24\times 10^{-1}$ and different charge values $q$. As $q$ increases, the radial tidal force tends to change from radial elongation to compression, with maxima shifted further away from the origin.

\subsection{Angular Tidal Forces}

\begin{figure*}[ht!]
\centering
\subfigure[$q = 0.3$] 
{\label{Fangulara}\includegraphics[width=8.5cm]{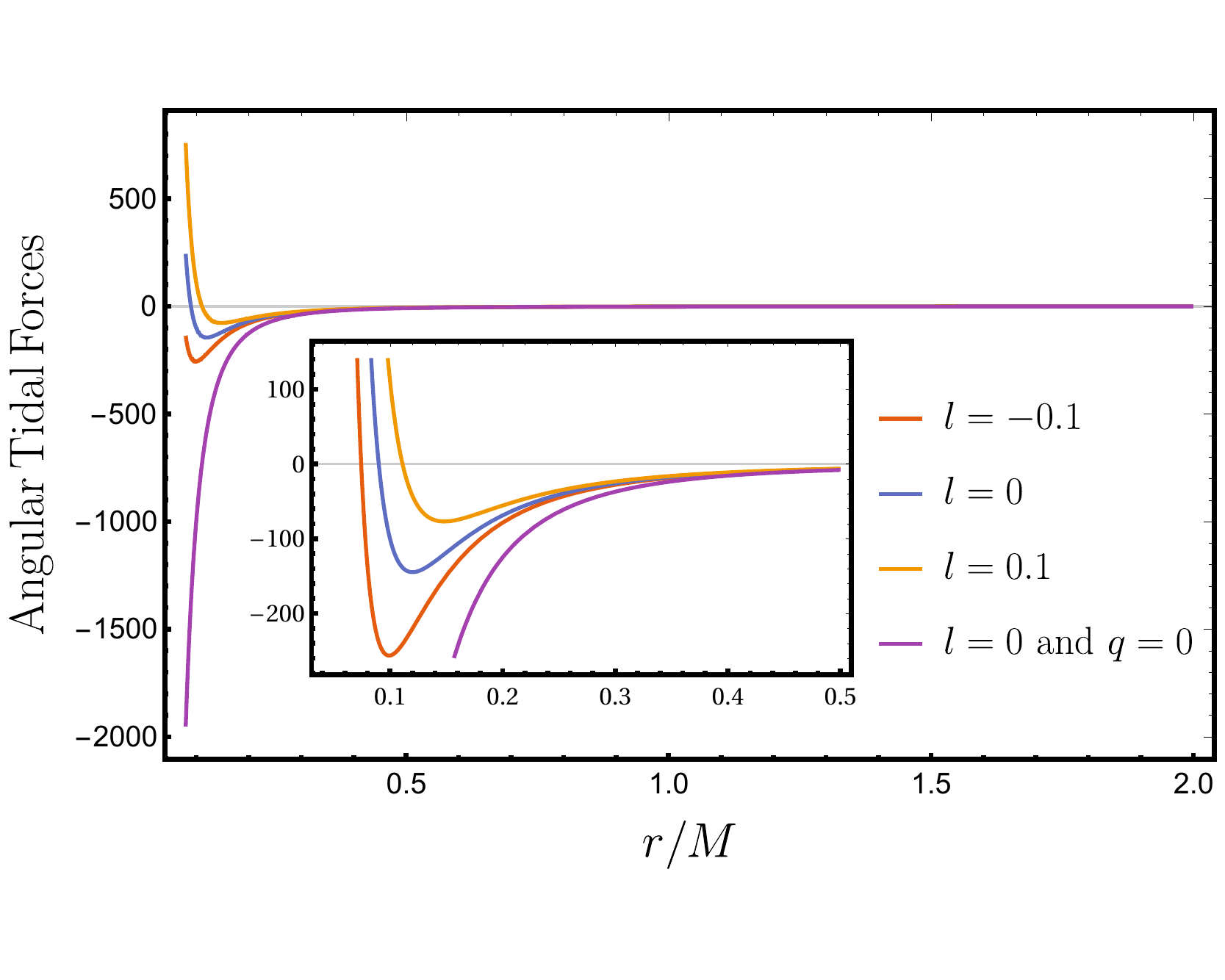} }
\hspace{0.1cm}
\subfigure[ $l=1.24\times 10^{-1}$] 
{\label{Fangularb}\includegraphics[width=8.5cm]{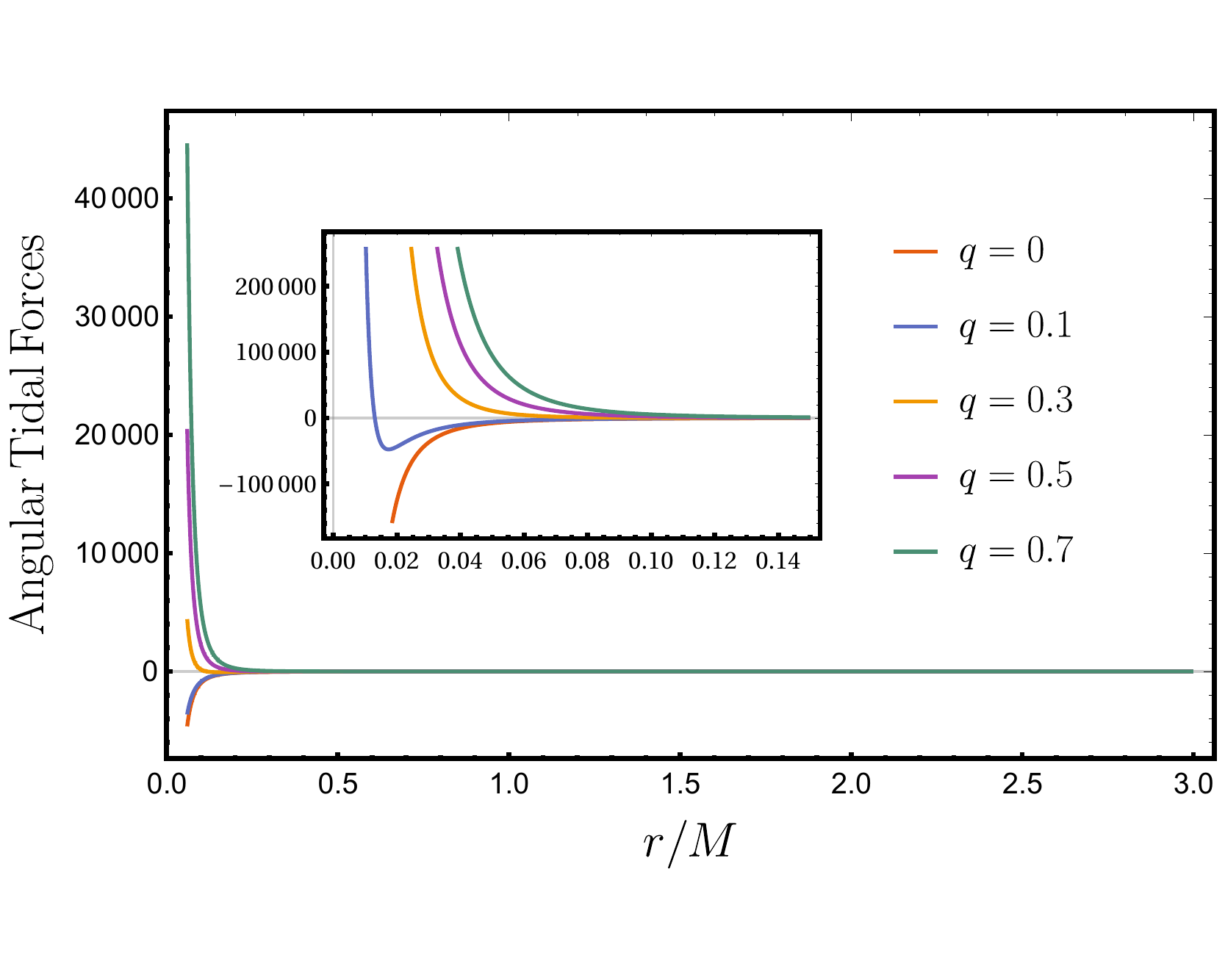} }
\caption{(a) Angular tidal forces with $q=0.3$ and different $l$.  (b) Angular tidal forces with $l=1.24\times 10^{-1}$ for different $q$. Note that we did not used negative values of $l$ because their effects can be neglected within the range obtained in \cite{KReletrico}.  }\label{Fig4}
\end{figure*}

\begin{figure*}[ht!]
\centering
\subfigure[$l = -4.59\times 10^{-3}$ (dashed line)] 
{\label{RARIO1}\includegraphics[width=8.5cm]{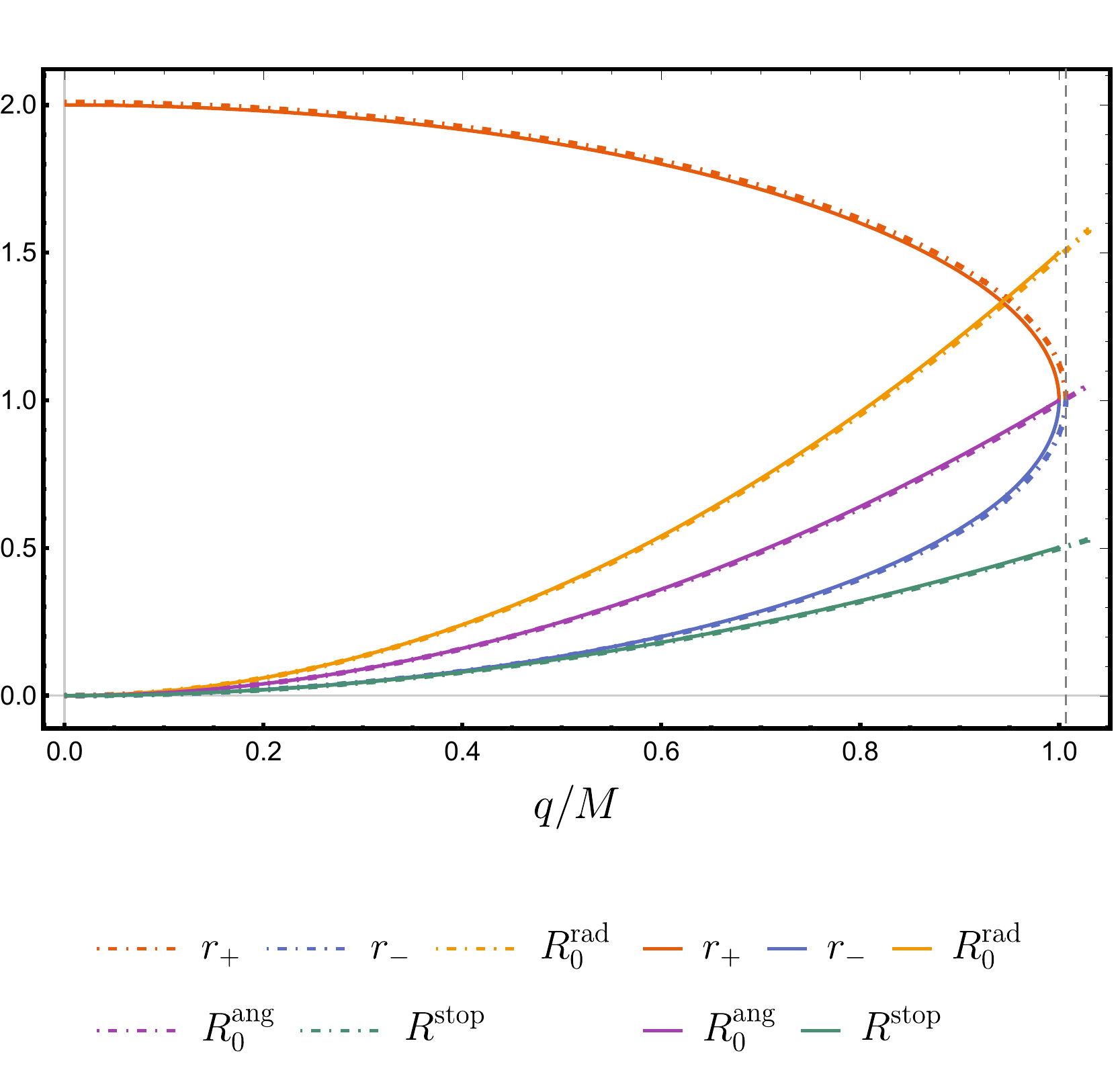} }
\hspace{0.1cm}
\subfigure[ $l=1.24\times 10^{-1}$ (dashed line)] 
{\label{RAIO2}\includegraphics[width=8.5cm]{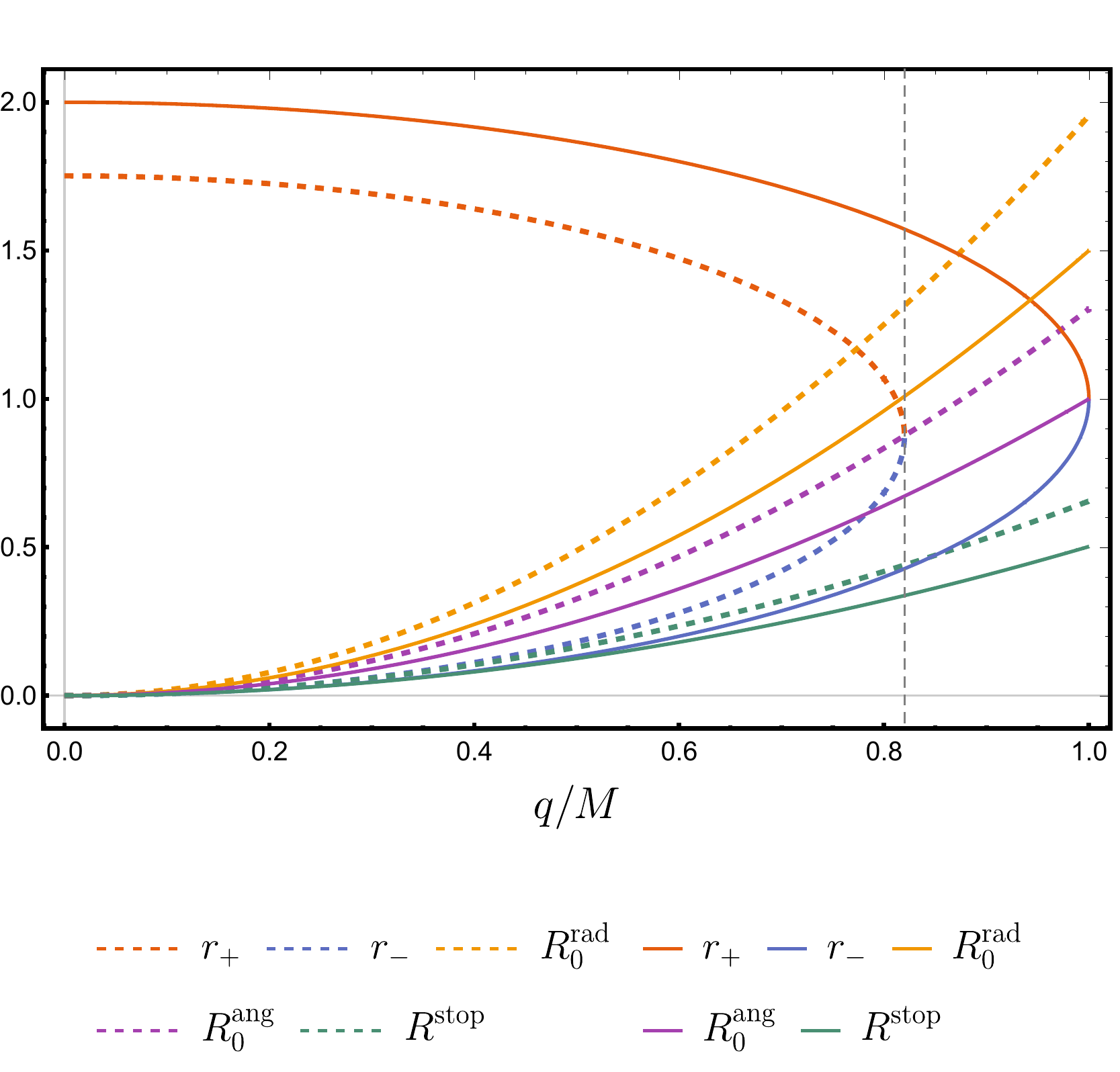} }
\caption{Event horizon $r_{\pm}$, $R^{\rm rad}_0$, $R^{\rm ang}_0$ and $R^{\rm stop}$ plotted in (a) for $l<0$ as functions of $q/M$. The vertical dashed line marks the position of the critical charge $q_c=1.00689 M$ for $l$. In (b) $r_{\pm}$, $R^{\rm rad}_0$, $R^{\rm ang}_0$ and $R^{\rm stop}$ are plotted for $l>0$ as functions of $q/M$ and the critical charge indicated by the vertical dashed line at $q_c= 0.819891 M$.}\label{Fig5}
\end{figure*}

Using Eqs.\,\eqref{KR0} and \eqref{angulari}, the angular tidal force in the charged KR black hole is
\begin{eqnarray}
\frac{D^2\eta^{\hat{i}}}{D\tau ^2}=\frac{1}{r^4}\left(\frac{q^2}{(1-l)^2}-M r\right)\eta^{\hat{i}}\,, \label{Fangular}
\end{eqnarray} 
which reduces to the expression found for the RN solution when $l\rightarrow 0$ and Schwarschild for $q=0$.  The angular tidal force is zero at 
\begin{eqnarray}
R^{\rm ang}_0=\frac{q^2}{(1-l)^2 M}\,,\label{angular0}
\end{eqnarray}
and minimum at
\begin{eqnarray}
R^{\rm ang}_{\rm min}=\frac{4 q^2}{3 (1-l)^2 M}\,. \label{angularmax}
\end{eqnarray} 

The behaviour of Eq.\,\eqref{Fangular} is plotted in Fig.\,\ref{Fangulara} for different values of $l$ with fixed $q=0.3$. We verify that the angular tidal force is zero at a single point given by Eq.\,\eqref{angular0}, shifted closer to the origin for $l<0$ and away from it when $l>0$. There is initially a compressive behaviour due to the angular tidal forces but, as the falling object reaches the point $R^{\rm ang}_{\rm min}$ it reverses its behaviour to stretching. The minimum of the force is shifted right as $l$ increases from $l<0$, leading the $R^{\rm ang}_{\rm min}$ further away from the origin. 

In Fig.\,\ref{Fangularb} we fix $l=1.24\times 10^{-1}$ and vary $q$. As $q$ increases, tidal forces tend to become stretching forces. However, for values very close to zero, these forces become of compressive nature up to a minimum value at  $R^{\rm ang}_{\rm min}$, that shifts to the right as $q$ increases. Note that for $q=0$ (Schwarzschild) the force is always compressing and diverges to (negative) infinity as the object approaches the origin (see the curve in red).

In Figs.\,\ref{RARIO1} and \ref{RAIO2}, we plot the behaviour of Eqs.\,\eqref{radial0}, \eqref{radialmax}, \eqref{angular0} and \eqref{angularmax}, beyond the horizons $r_{\pm}$, for $l=-4.59\times 10^{-3}$ and $l=1.24\times 10^{-1}$ obtained in \cite{KReletrico}, and which are compared with the case $l=0$. We see that the values of $l$ obtained from the constraints imposed by the EHT observations change the scale size of the horizons, showing that they are also shifted as $l$ changes: shrinking for $l<0$ and enlarging for $l>0$. The inflection point $R^{\rm stop}$  is the point where the test object resumes its radial motion.   Note that $R^{\rm ang}_0$ is located inside the event horizon and outside the Cauchy horizon for all values of $l$. The intersection of $r_+$ with $R^{\rm rad}_0$, where the radial tidal force reverses its direction and becomes compressive, occurs at $q/M=\frac{2}{3} \sqrt{2} (1-l)^{3/2}$. Note also that the critical charge is given by $q_c/M=\sqrt{(1-l)^3}$; for instance, for  $l=-4.59\times 10^{-3}$ have  $q_c/M=1.00689$ and for $l=1.24\times 10^{-1}$, $q_c/M=0.819891$. In such a case the black hole becomes extremal with a single horizon,  marked by dashed gray lines in Fig.\,\ref{Fig5}.

Note that the metric given by   Eq.\,\eqref{KR0} with $q=0$ boils down to the uncharged KR metric \cite{KR}. However, the behavior of the radial and angular tidal forces for this case are identical to the Shchwarzschild solution, which can be seen from Eqs.\,\eqref{radial1} and \eqref{angulari} and depend on the derivatives of $A(r)$. This is an important result for analyzing the behaviour of objects close to the uncharged KR-BH and shows that the spontaneous Lorentz symmetry-breaking parameter has no effect on the dynamics of free fall in this space-time.

\section{Solving the geodesic deviation equations}\label{sec5}

Equations \eqref{radial1} and \eqref{angulari} can be solved to find the geodesic deviation vectors of a free-falling test object in the charged KR black hole as a function of the coordinate $r$.  Rewriting Eq.\,(\ref{rponto}) as
\begin{eqnarray}
\frac{dr}{d\tau}=-\sqrt{E^2-A(r)}\,,\label{Eb}
\end{eqnarray}
and using Eqs.\,\eqref{radial1} and \eqref{angulari}, we obtain
\begin{eqnarray}
\left[E^2-A(r)\right]\frac{d^2\eta^{\hat{1}}}{dr^2}-\frac{A'(r)}{2}\frac{d\eta^{\hat{1}}}{dr}+\frac{A''(r)}{2}\eta^{\hat{1}}&=&0\,,\\\label{eqdesv1}
\left[E^2-A(r)\right]\frac{d^2\eta^{\hat{i}}}{dr^2}-\frac{A'(r)}{2}\frac{d\eta^{\hat{i}}}{dr}+\frac{A'(r)}{2r}\eta^{\hat{i}}&=&0\,.
\label{eqdesvi}
\end{eqnarray}
These yield analytical solutions for the radial  $\eta^{\hat{1}}$ and angular $\eta^{\hat{i}}$ geodesic deviation components, which are given by the expressions
\begin{eqnarray}
\eta^{\hat{1}}&=&\sqrt{E^2-A(r)}\left[B_1+B_2\int\frac{dr}{\left(E^2-A(r)\right)^{3/2}}\right],\label{desvio11}\\
\eta^{\hat{^i}}&=&\left[B_3+B_4\int\frac{dr}{r^2\left(E^2-A(r)\right)^{1/2}}\right]r\,,\label{desvioi}
\end{eqnarray}
respectively, where $B_1$, $B_2$, $B_3$ and $B_4$ are constants of integration. For an object released from rest at $r=b>r_+$ towards the charged KR black hole, using Eq.\,\eqref{Eb} these constants read as
\begin{eqnarray}
B_1&=&\frac{b^3 (1-l)^2}{b (1-l)^2 M-q^2}\frac{d\eta^{\hat{1}}(b)}{d\tau}\,,\label{B1}\\B_2&=& \frac{1}{b^3} \left(b M-\frac{q^2}{(1-l)^2}\right) \eta^{\hat{1}}(b)\,,\label{B2} \\
B_3&=&\frac{1}{b}\eta^{\hat{i}}(b)\,\,\,\text{and }\,\,B_4=-b \frac{d\eta^{\hat{i}}(b)}{d\tau}\,, \label{B3B4}
\end{eqnarray}
where the last set of equations \eqref{B3B4} are identical to those of the Schwarzschild solution \cite{Crispino, holographictidal}.  The quantities $\eta^{\hat{1}}(b)$ and $\eta^{\hat{i}}(b)$,  $\frac{d\eta^{\hat{1}}(b)}{d\tau}$ and $\frac{d\eta^{\hat{i}}(b)}{d\tau}$ are, respectively, the initial deviations and initial speeds between two geodesics at $r=b$, in the radial and angular directions.   
Therefore, the radial component of the geodesic deviation \eqref{desvio11} is given by
\begin{eqnarray}
&\eta^{\hat{1}}=\eta^{\hat{1}}(b) \Bigg[ g(r)
- h(r)\sinh ^{-1}\left(\frac{\sqrt{b-r} \sqrt{2 b (1-l)^2 M-q^2}}{\sqrt{2b}  \sqrt{q^2-b (1-l)^2 M}}\right)\Bigg]
\nonumber\\
&
+\frac{b^3 (1-l)^2}{b (1-l)^2 M-q^2}\frac{d\eta^{\hat{1}}(b)}{d\tau}
\sqrt{\frac{(r-b) \left(b \left(q^2-2 (1-l)^2 M r\right)+q^2 r\right)}{b^2 (1-l)^2 r^2}}\,,\label{Eta1}
\end{eqnarray}
%
%
where the functions $g(r)$ and $h(r)$ read explicitly
\begin{eqnarray}
g(r)&=&\frac{1}{b r \left(b (1-l)^2 M-q^2\right) \left(q^2-2 b (1-l)^2 M\right)^2}\nonumber\\
&&\qquad \times \Bigg[b^2 (1-l)^2 M r \Big(6 b^2 (1-l)^4 M^2\nonumber\\
&&-10 b (1-l)^2 M q^2+5 q^4\Big)+b^2 q^2 \big(-3 b^2 (1-l)^4 M^2\nonumber\\
&&+4 b (1-l)^2 M q^2-2 q^4\big)+r^2 \left(q^2-b (1-l)^2 M\right)^2\nonumber\\
&&\qquad \times\left(q^2-2 b (1-l)^2 M\right)\Bigg]
\end{eqnarray}
and 
\begin{eqnarray}
h(r)&=&\frac{6 b^{3/2} (1-l)^2 M\sqrt{b-r}}{r \left(2 b (1-l)^2 M-q^2\right)^{5/2}}\left(q^2-b (1-l)^2 M\right)^{3/2} \nonumber\\
&&\times\sqrt{\frac{b \left(q^2-2 (1-l)^2 M r\right)+q^2 r}{q^2-b \left(b (1-l)^2 M\right)}}\,,
\end{eqnarray}
respectively. 

As for the angular component of the geodesic deviation, it reads as
\begin{eqnarray}
\eta^{\hat{i}}&=&r \Bigg[\frac{1}{b}\eta^{\hat{i}}(b)+\frac{2 b  (1-l)}{q}\frac{d\eta^{\hat{i}}(b)}{d\tau}\nonumber\\ &&\times\tan ^{-1}\left(\frac{q \sqrt{b-r}}{\sqrt{2 b (1-l)^2 M r-b q^2-q^2 r}}\right)\Bigg] \,.
\end{eqnarray}

All the above expressions recover the results for the case of RN for $l \rightarrow 0$, that is,
\begin{eqnarray}
&\eta^{\hat{1}}=\eta^{\hat{1}}(b) \Bigg[ \mathcal{G}(r)
- \mathcal{H}(r)\sinh ^{-1}\left(\frac{\sqrt{b-r} \sqrt{2 b  M-q^2}}{\sqrt{2b}  \sqrt{q^2-b  M}}\right)\Bigg]
\nonumber\\
&
+\frac{b^3 }{b  M-q^2}\frac{d\eta^{\hat{1}}(b)}{d\tau}
\sqrt{\frac{(r-b) \left(b \left(q^2-2  M r\right)+q^2 r\right)}{b^2  r^2}}\,,
\end{eqnarray}
where the functions $\mathcal{G}(r)$ and $\mathcal{H}(r)$ read explicitly
\begin{eqnarray}
\mathcal{G}(r)&&=\frac{1}{b r \left(b  M-q^2\right) \left(q^2-2 b  M\right)^2} \Bigg[b^2  M r \Big(6 b^2  M^2\nonumber\\
&&-10 b  M q^2+5 q^4\Big)+b^2 q^2 \big(-3 b^2  M^2+4 b  M q^2-2 q^4\big)\nonumber\\
&&+r^2 \left(q^2-b  M\right)^2
\left(q^2-2 b  M\right)\Bigg]
\end{eqnarray}
and 
\begin{eqnarray}
\mathcal{H}(r)&=&\frac{6 b^{3/2}  M\sqrt{b-r}}{r \left(2 b  M-q^2\right)^{5/2}}\left(q^2-b  M\right)^{3/2} \nonumber\\
&&\times\sqrt{\frac{b \left(q^2-2  M r\right)+q^2 r}{q^2-b \left(b  M\right)}}\,,
\end{eqnarray}
and the angular component
\begin{eqnarray}
\eta^{\hat{i}}&=&r \Bigg[\frac{1}{b}\eta^{\hat{i}}(b)+\frac{2 b }{q}\frac{d\eta^{\hat{i}}(b)}{d\tau}\nonumber\\ &&\times\tan ^{-1}\left(\frac{q \sqrt{b-r}}{\sqrt{2 b  M r-b q^2-q^2 r}}\right)\Bigg] \,.
\end{eqnarray}
In the limit that $q\rightarrow 0$ and $l\rightarrow 0$, we recove Schwarzschild,
\begin{eqnarray}
\eta^{\hat{1}}(r)=&&\sqrt{\frac{2bM(b-r)}{r}}\frac{1}{M}\frac{d\eta^{\hat{1}}(b)}{d\tau}+\frac{1}{2}\eta^{\hat{1}}(b)\left(3-\frac{r}{b}\right)\nonumber\\
&&+\frac{3}{2}\sqrt{\frac{b}{r}-1}\,\arccos\left[\left(\frac{r}{b}\right)^{1/2}\right]\eta^{\hat{1}}(b)\,,
\end{eqnarray}
\begin{eqnarray}
&&\eta^{\hat{i}}(r)=\left[\frac{1}{b}\eta^{\hat{i}}(b)+\frac{d\eta^{\hat{i}}(b)}{d\tau}\sqrt{\frac{2b}{M}\left(\frac{b}{r}-1\right)}\right]r\,.
\end{eqnarray}
all of which correspond to the expressions obtained in \cite{Crispino}.

\begin{figure*}[t!]
\centering
\subfigure[$q = 0.6 $ ] 
{\label{Figdesva1}\includegraphics[width=8.5cm]{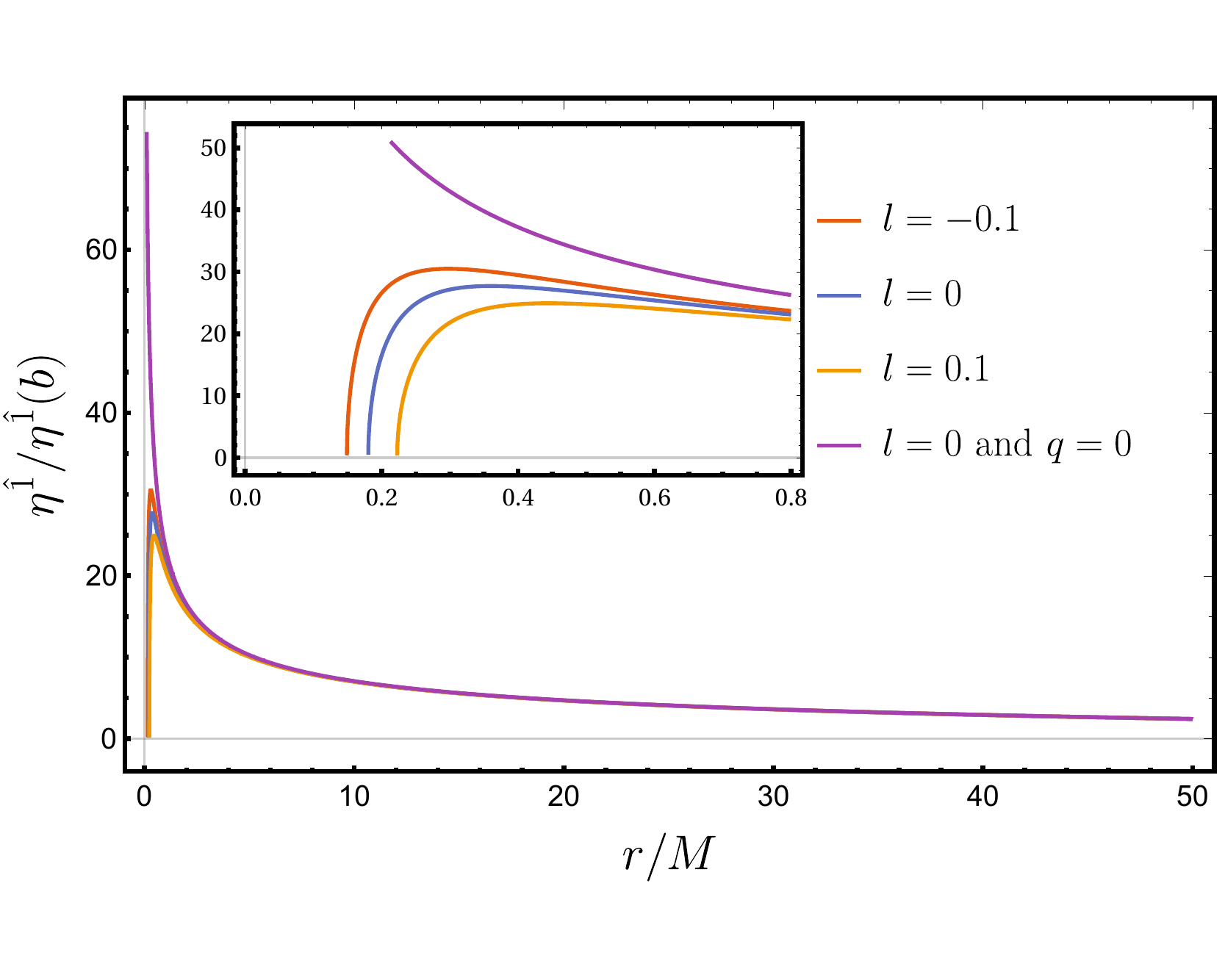} }
\subfigure[ $l=1.24\times 10^{-1}$ ] 
{\label{Figdesvb1}\includegraphics[width=8.5cm]{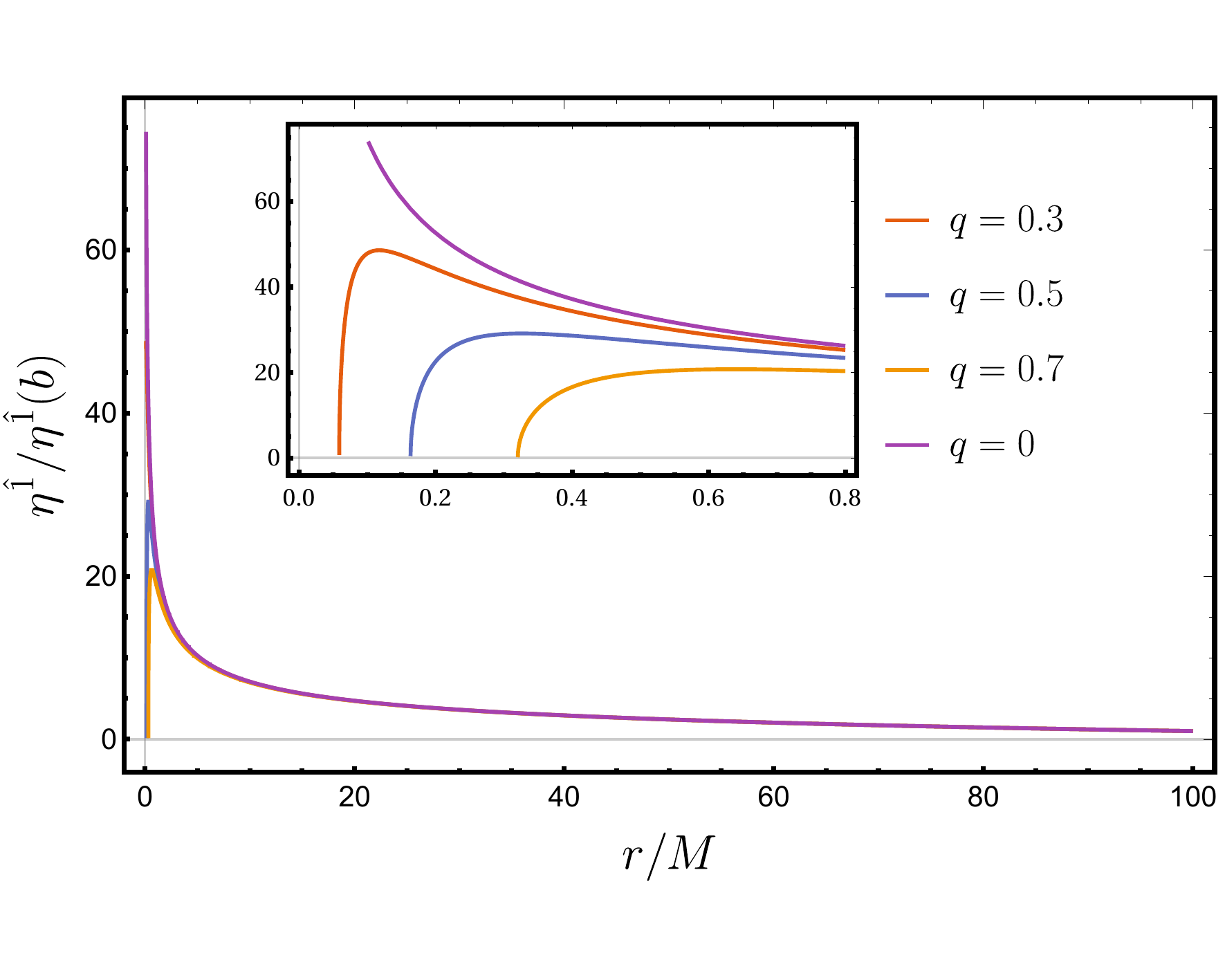} }
\caption{In (a) we have plotted the radial component of the geodesic deviation for different values of $l$ and fixed $q$, In (b) we have plotted the radial geodesic deviation for different values of $q$ and $l$ fixed. Both of them for condition $\mathcal{CI}$.}\label{Fig6}
\end{figure*}

To analyze the behaviour of the geodesic deviation functions, consider two types of initial conditions.  First, for the object starting from rest at $r=b$, we have the initial conditions $\mathcal{CI}=\lbrace\eta^{\hat{1}}(b)>0\,,\, d\eta^{\hat{1}}(b)/d\tau=0\,, \, \eta^{\hat{i}}>0\,,\, d\eta^{\hat{i}}(b)/d\tau=0\rbrace$. 
Fig.\,\ref{Fig6} shows the effect of the spontaneous Lorentz symmetry-breaking parameter on the geodesic deviation for the radial component of the force for these initial conditions with different values of $l$ (Fig.\,\ref{Figdesva1}), and also with fixed $l$ for different values of $q$ (Fig.\,\ref{Figdesvb1}). 

\begin{figure}[t!]
\centering
{\includegraphics[width=8.5cm]{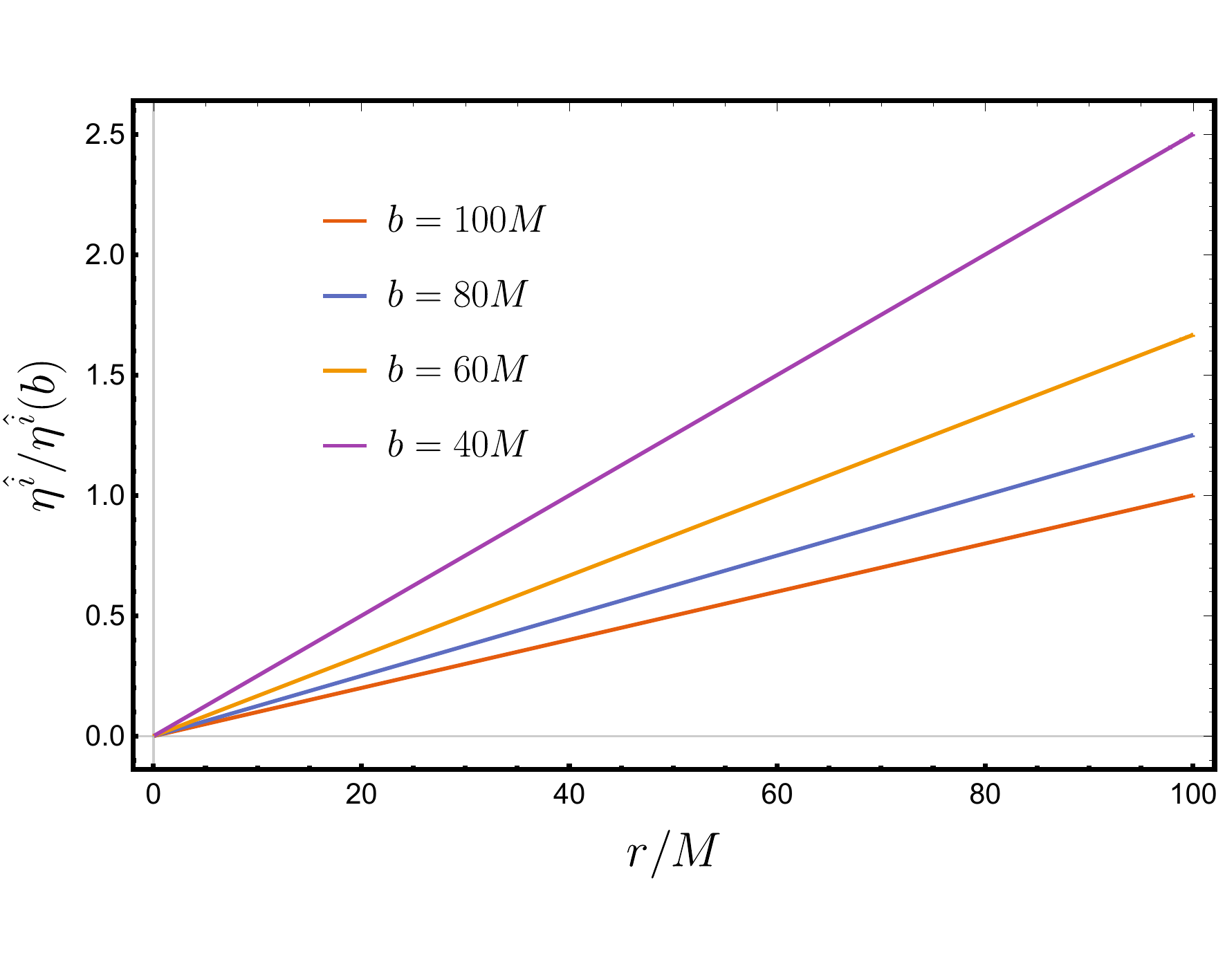} }
\caption{Angular component of the geodesic deviation with different initial values $b$ and assuming $\mathcal{CI}$. }\label{Fig7}
\end{figure}

Note that in Fig.\,\ref{Figdesva1} for different $l$, the radial component of the geodesic deviation reaches a limiting value, then decreases as the object falls towards the inside horizon $r_-$. The limit value of the component decreases as $l$ increases, similar to the effect of charge in the RN case. This effect is the opposite of the one observed for the Schwarzschild solution, which crosses the event horizon and tends to infinity due to the radial tidal force of elongation, which is also infinite. In Fig.\,\ref{Figdesvb1}, we fix $l$ and vary $q$, finding that the result is similar to that found previously for RN \cite{Crispino}. The angular component for the condition $\mathcal{CI}$ is depicted in Fig.\,\ref{Fig7} for different values of $b$, which decreases linearly with $r$ to zero as in the RN and Schwarzschild solutions. 

\begin{figure*}[t!]
\centering
\subfigure[$q = 0.6 $ ] 
{\label{Figdesva2}\includegraphics[width=8.5cm]{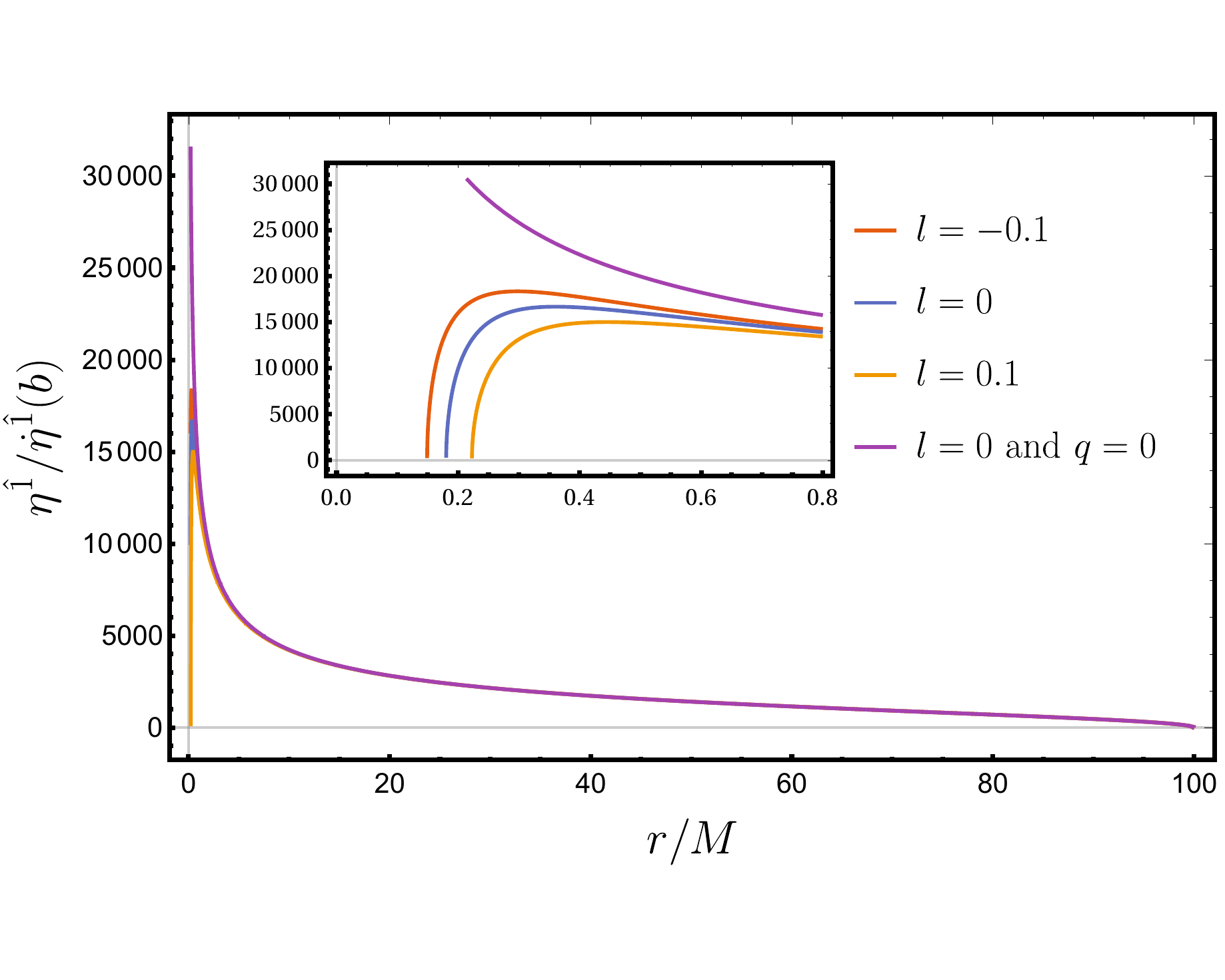} }
\hspace{0.1cm}
\subfigure[ $l=1.24\times 10^{-1}$ ] 
{\label{Figdesvb2}\includegraphics[width=8.5cm]{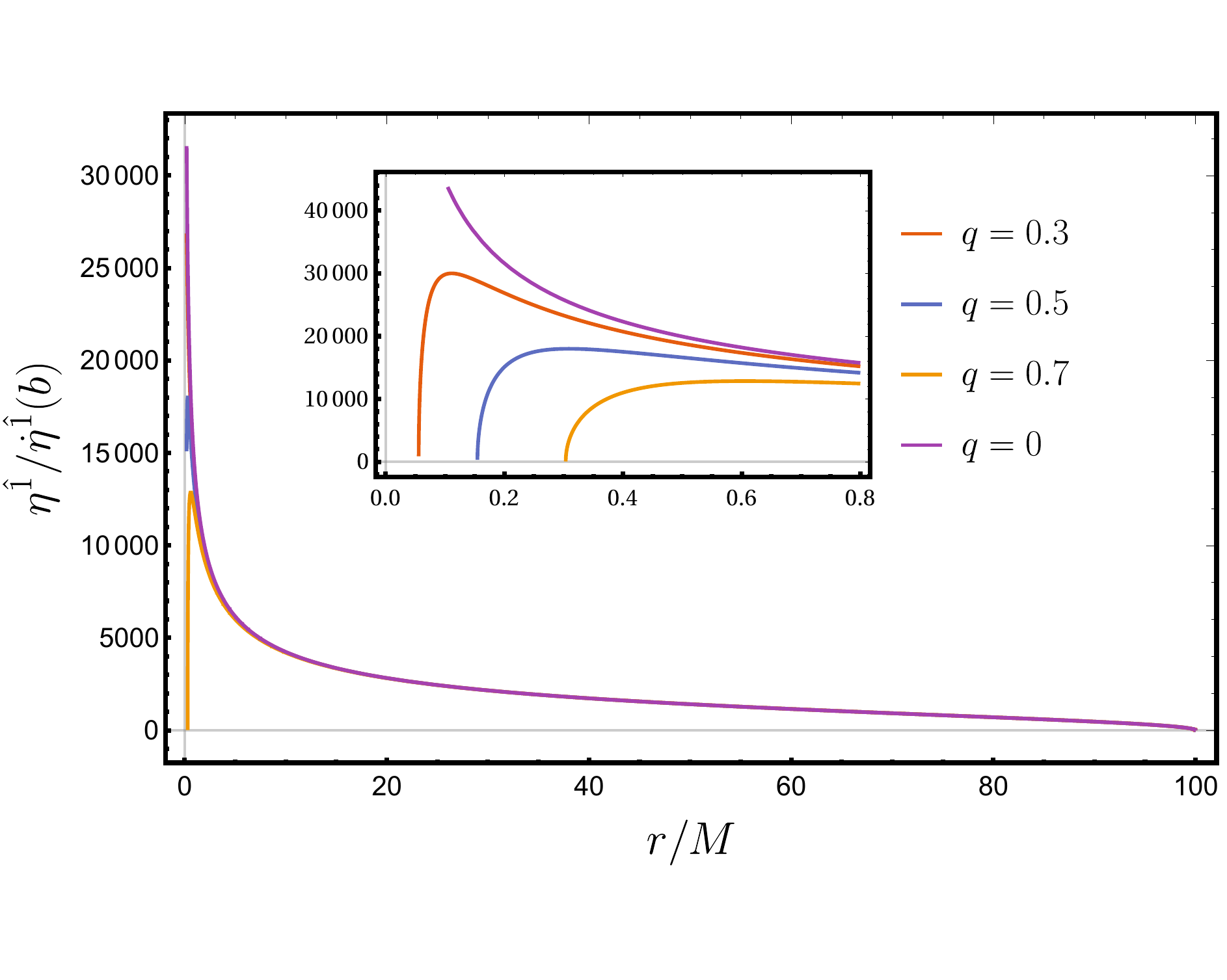} }
\caption{In (a) we have plotted the radial component of the geodesic deviation for different values of $l$ and fixed $q$, plot (a).  In (b) we have plotted radial geodesic deviation for different values of $q$ and fixed $l$. Both of them for condition $\mathcal{CII}$.}\label{Fig8}
\end{figure*}

\begin{figure*}[t!]
\centering
\subfigure[$q = 0.6 $ ] 
{\label{Figdesva3}\includegraphics[width=8.5cm]{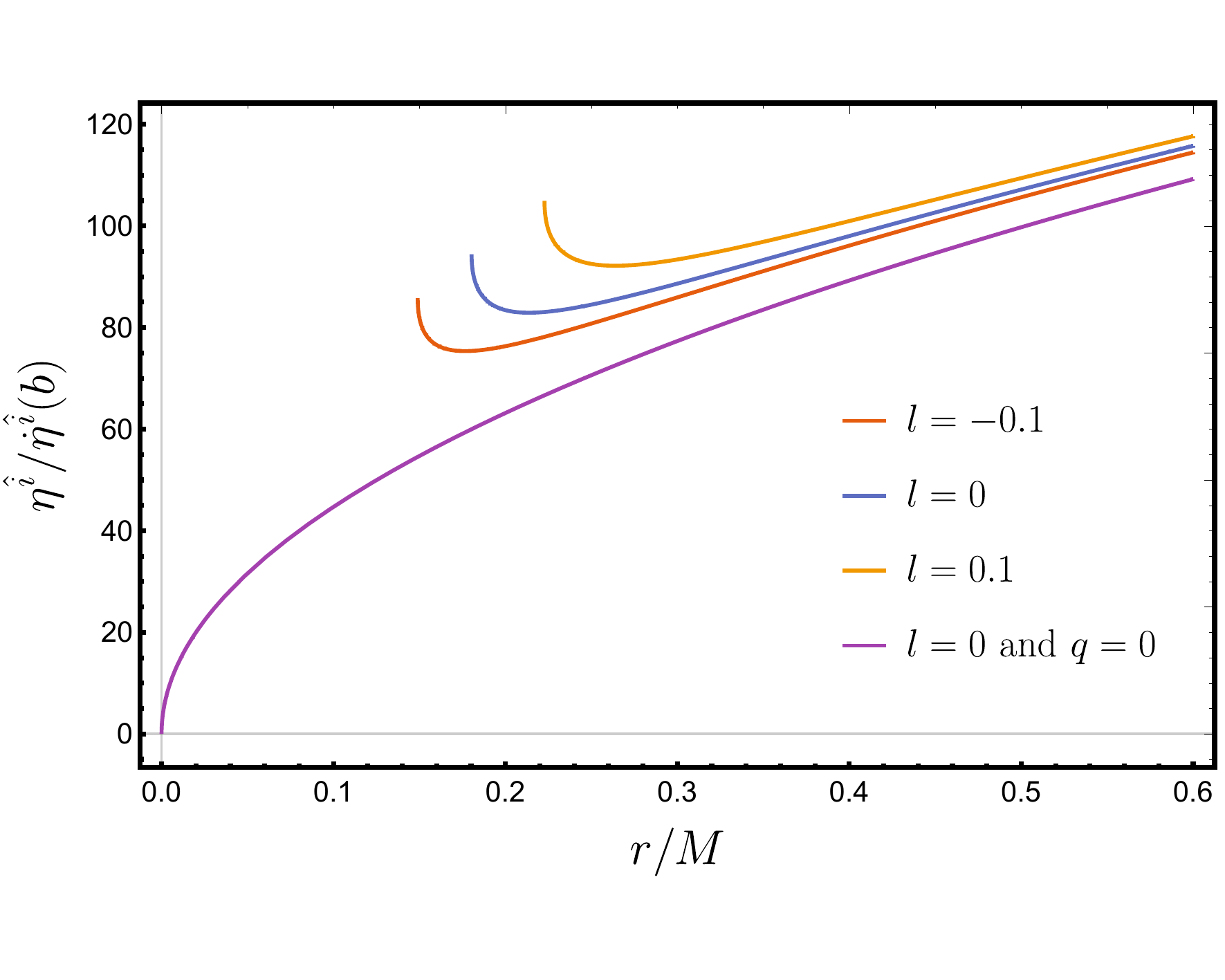} }
\subfigure[ $l=1.24\times 10^{-1}$ ] 
{\label{Figdesvb3}\includegraphics[width=8.5cm]{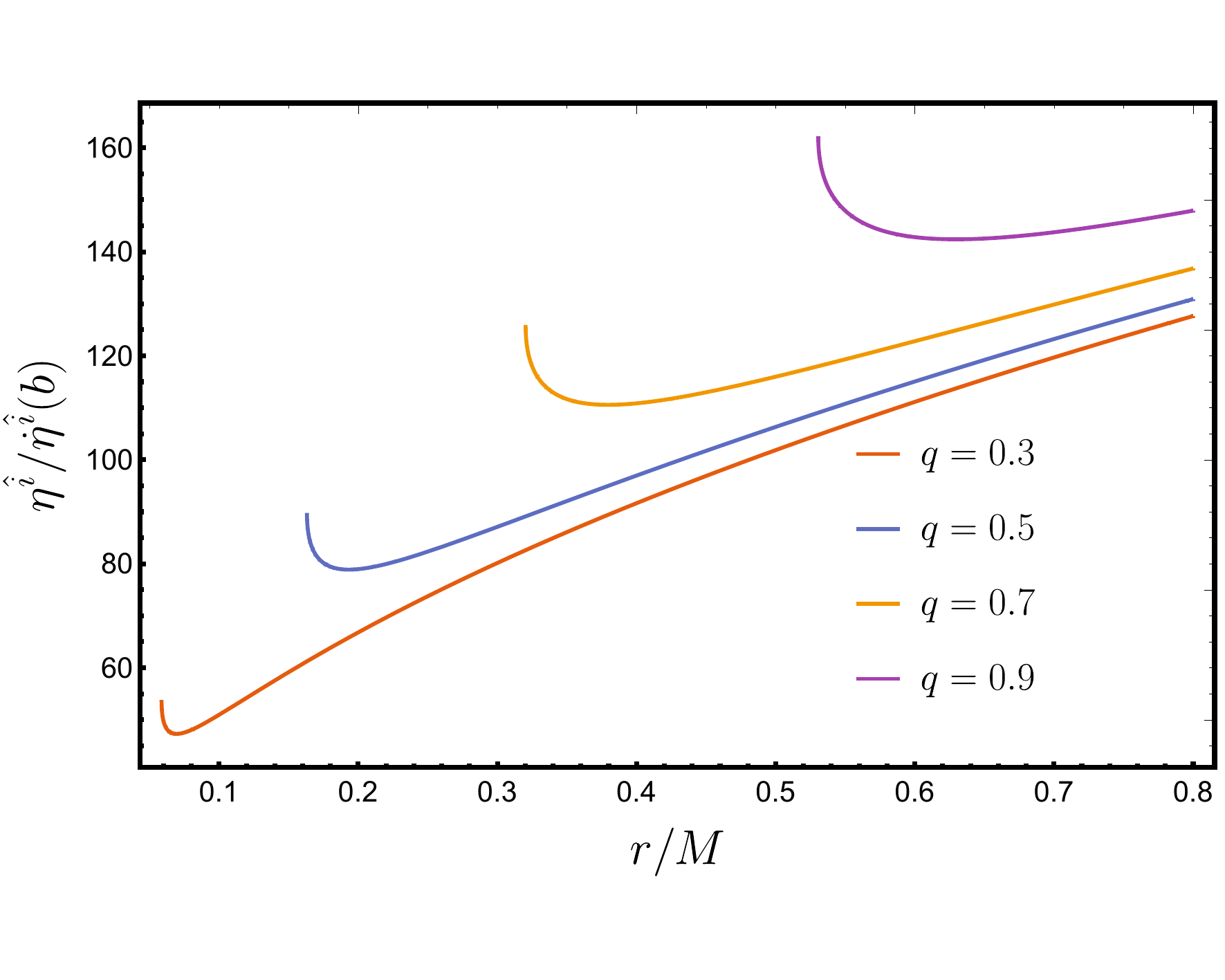} }
\subfigure[$l=1.24\times 10^{-1}$ and $q=0.6$] 
{\label{Figdesvc3}\includegraphics[width=8.5cm]{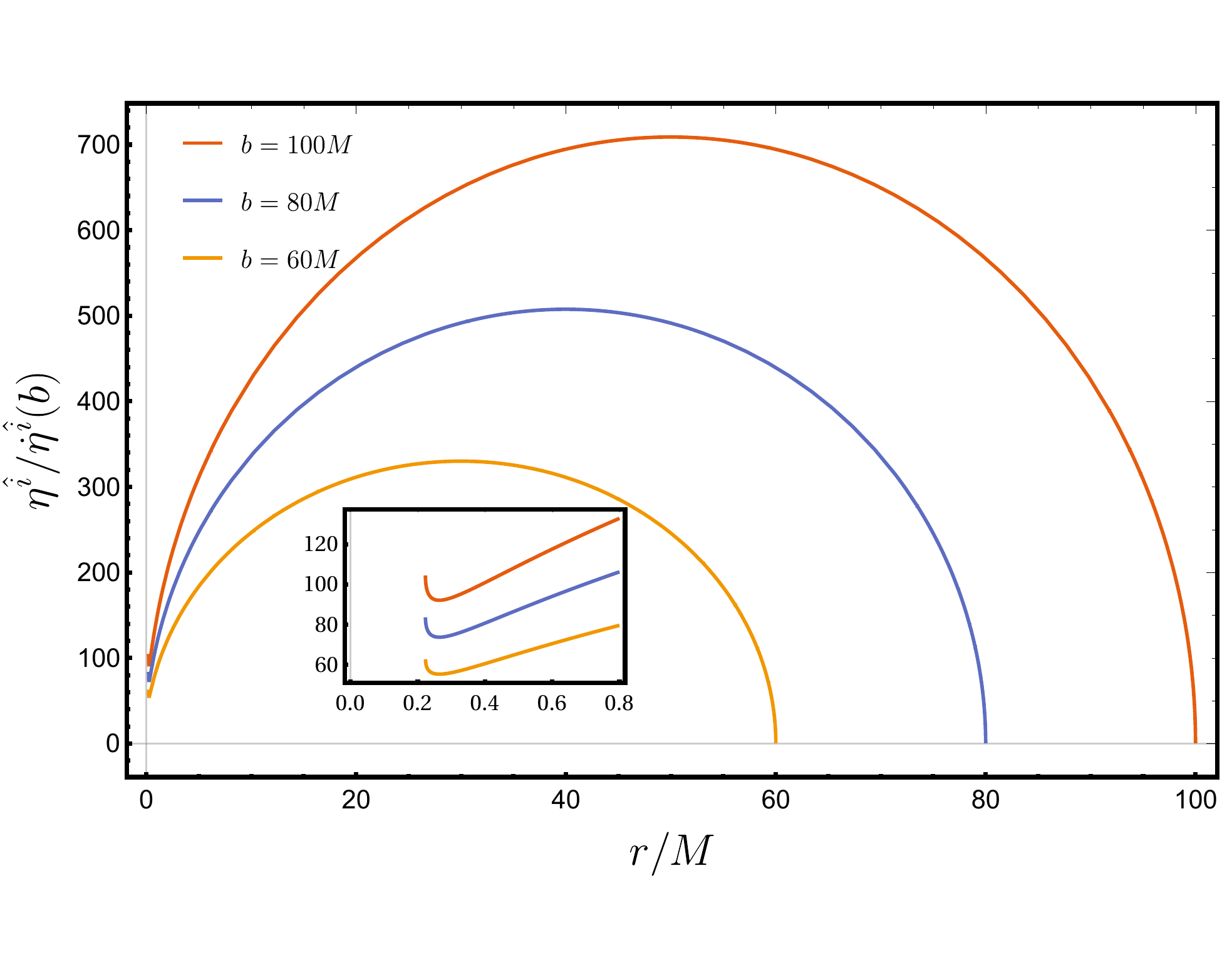}}
\caption{Angular component of the geodesic deviation for different values of $l$ and fixed $q$ (in addition to $q=0$ for comparison with Schwarzschild), plot (a).  In (b) we have the plot of the angular geodesic deviation for different values of $q$ and fixed $l$. In (c) we vary $b$ and fix $q$ and $l$.  Both for condition  $\mathcal{CII}$.}\label{Fig9}
\end{figure*}

We consider a second initial condition given by $\mathcal{CII}=\lbrace\eta^{\hat{1}}(b)=0\,,\, d\eta^{\hat{1}}(b)/d\tau>0\,, \, \eta^{\hat{i}}(b)=0\,,\, d\eta^{\hat{i}}(b)/d\tau>0\rbrace$. 
Fig.\,\ref{Fig8} depicts the radial component for $\mathcal{CII}$, which corresponds to a similar behavior as condition $\mathcal{CI}$. The angular component is plotted in Fig.\,\ref{Figdesva3}. We see an initial increase to a maximum at $b/2$ from where it starts to decrease. This is due to compression from tidal forces. Note that the effect of $l$ on the angular geodesic deviation for large $r$ is subtle, however, near $r=0$, the effects of $l$ significantly differ from the case $l=0$ and $q=0$.  After reaching a minimum, the angular component begins to increase until it reaches $r=R^{\rm stop}$. This behavior reflects the change in sign of the angular component. By fixing the value of $l$ and adopting different values of $q$, the analog to RN is recovered, as can be seen in Fig.\,\ref{Figdesvb3}. For Schwarzschild the behaviour is similar, except near $r=0$ where it continues to decrease to zero after passing through the maximum point. This reflects the infinite tidal force near the singularity.  Fig.\,\ref{Figdesvc3} shows that for the considered values of $l$ and $q$, as $r$ decreases, the angular geodesic deviation also decreases.  

\section{Other metrics with spontaneously Lorentz symmetry-breaking}\label{sec6}
\subsection{Lorentz symmetry-breaking RN-like} 
To compare the results with those corresponding to other theories implementing a Lorentz symmetry-breaking, let us analyze the tidal effects of objects whose space-time was also obtained by coupling the KR field, as obtained in  \cite{KRMaluf}
\begin{eqnarray}
A(r)=1-\frac{R_s}{r}+\frac{\Upsilon}{r^{2/\lambda}}\,,\label{AKR2}
\end{eqnarray}
where $\lambda:=|b|^2\xi_2$ and $\Upsilon$ are parameters that determine the effects of the Lorentz violation and ${R_s}=2GM$ is Schwarzschild radius. Since the KR pseudoelectric field is constant, this metric recovers the same structure of the RN space-time when $\lambda\rightarrow 1$. However, Schwarzschild is fully recovered when $\Upsilon\rightarrow 0$. In \cite{KRlambda}, the constraints for the two parameters, $\lambda$ and $\Upsilon$, were obtained from the angular diameter of the shadow of SgrA*. We next evaluate the tidal effects for this space-time.

\begin{figure*}[ht!]
\centering
\subfigure[$\Upsilon = 0.2$] 
{\label{radlam1}\includegraphics[width=8.5cm]{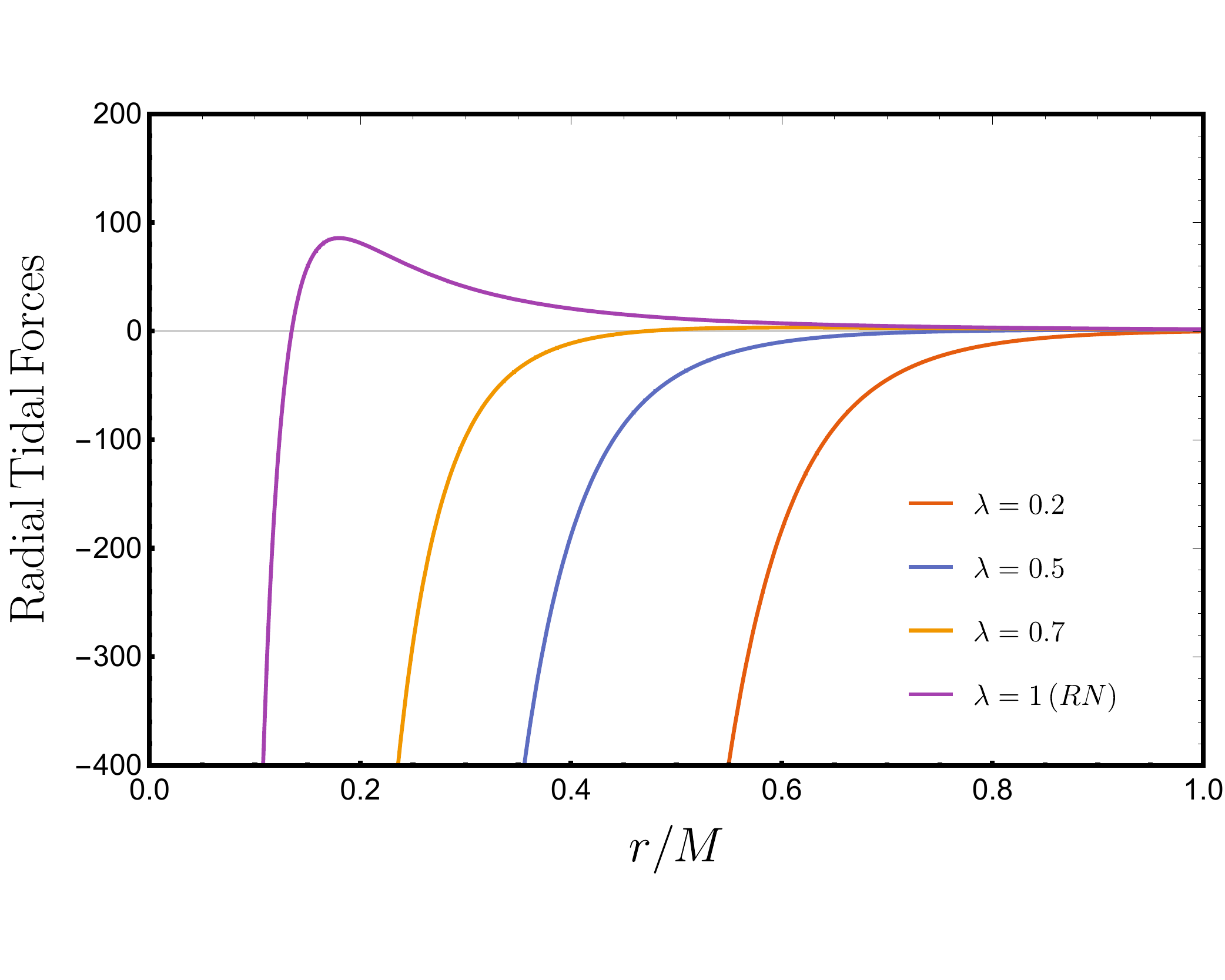} }
\hspace{0.1cm}
\subfigure[ $\lambda=0.6$] 
{\label{radup1}\includegraphics[width=8.5cm]{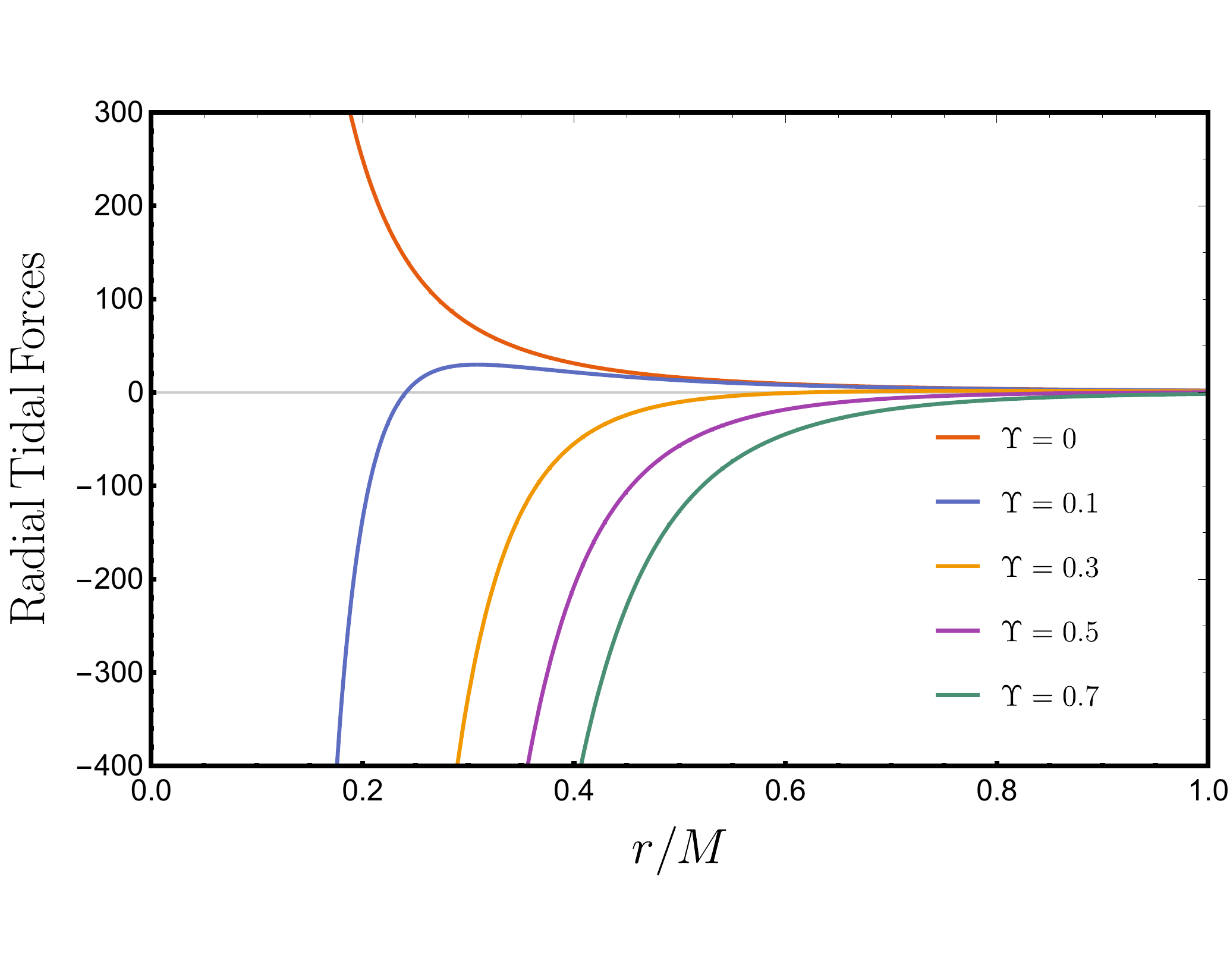} }
\subfigure[$\Upsilon = 0.2$] 
{\label{anglam1}\includegraphics[width=8.5cm]{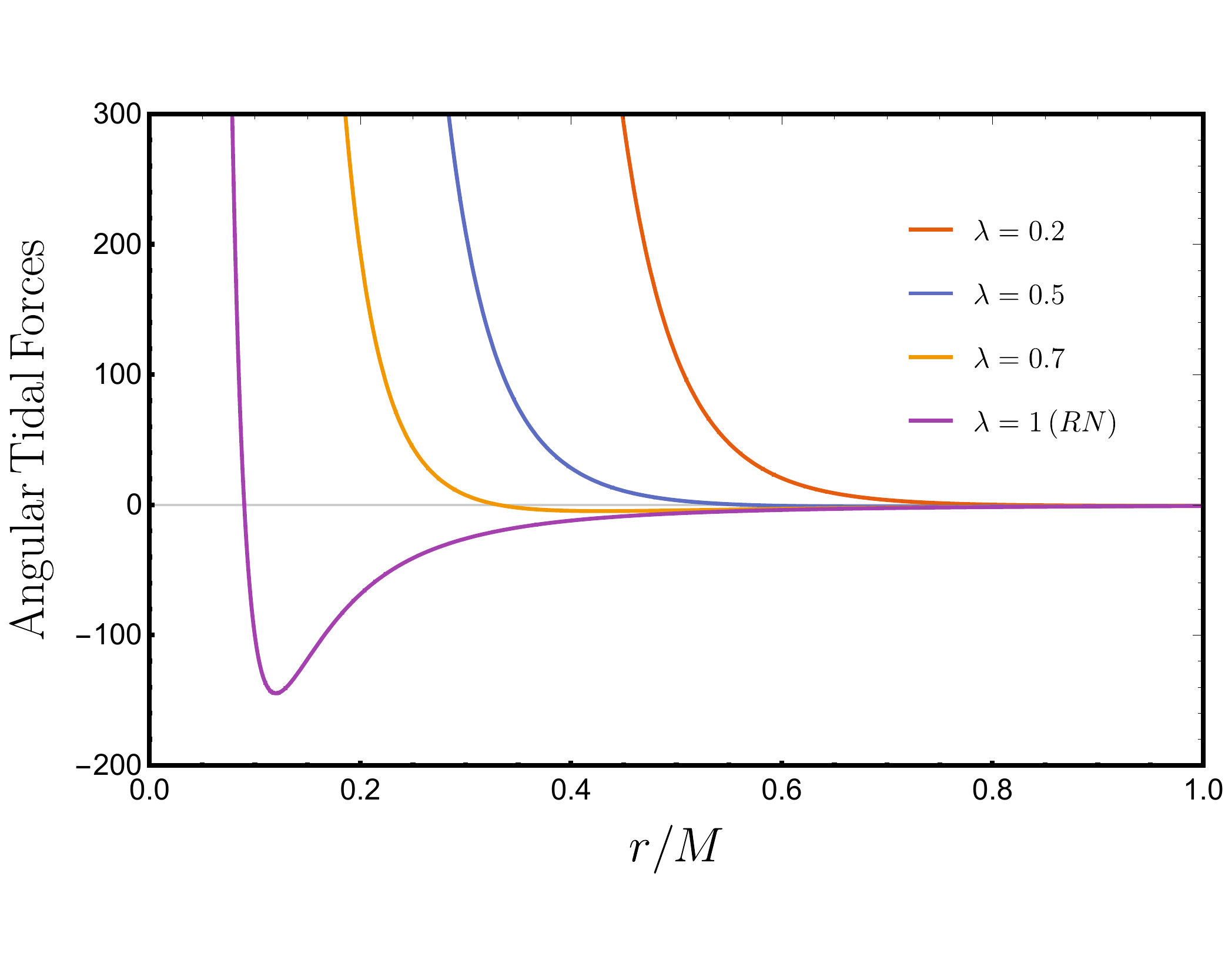}}
\hspace{0.1cm}
\subfigure[$\lambda=0.6$] 
{\label{angup1}\includegraphics[width=8.5cm]{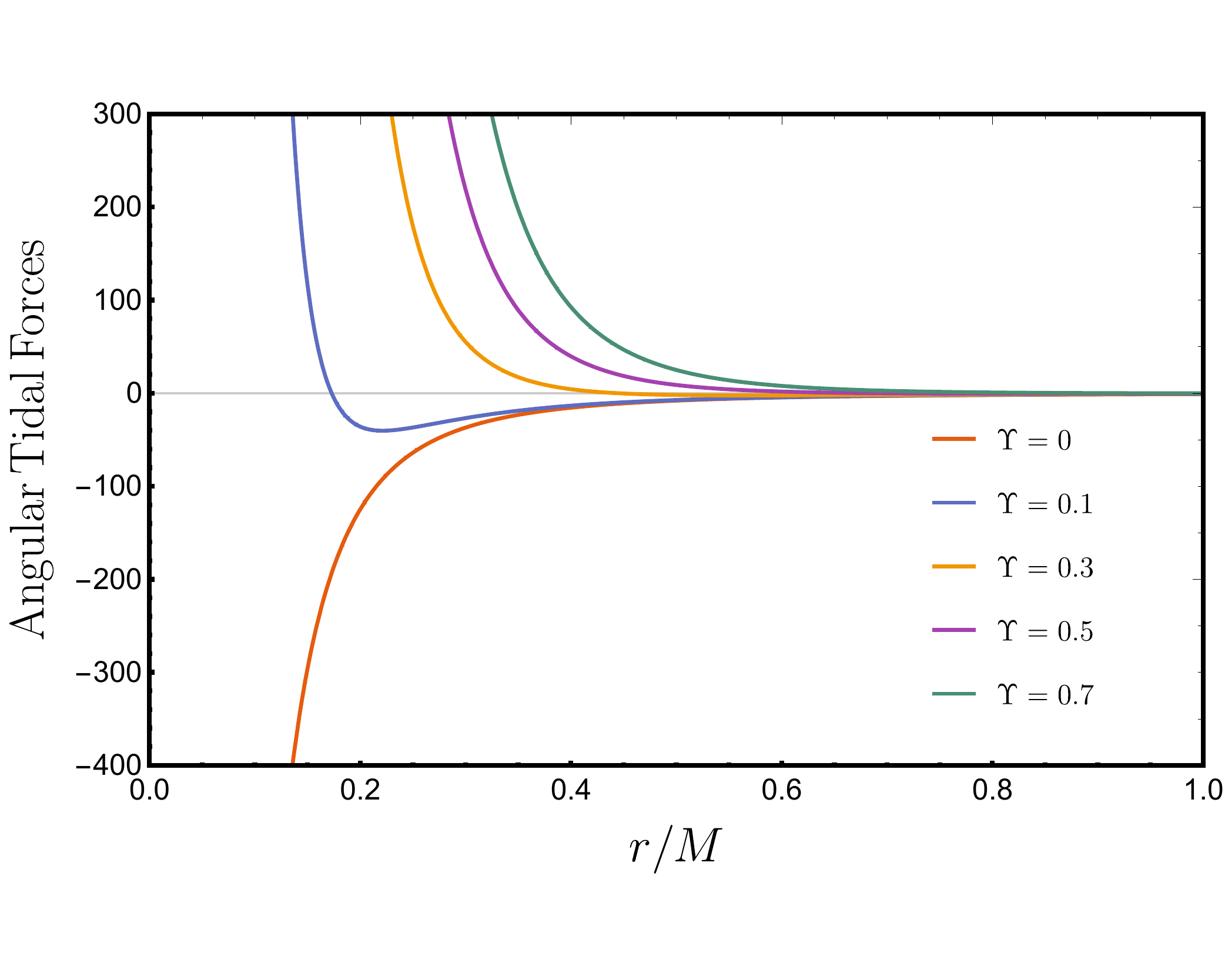}}
\caption{(a) Radial tidal forces with $\Upsilon=0.2$ and different $\lambda$.  (b) Radial tidal forces with $\lambda=0.6$ for different $\Upsilon$. (c) Angular tidal forces with $\Upsilon=0.2$ and different $\lambda$.  (c) Angular tidal forces with $\lambda=0.6$ for different $\Upsilon$.}\label{Fig10}
\end{figure*}

Using Eqs.\eqref{radial1} and \eqref{angulari} we obtain the radial and angular components for tidal force
\begin{eqnarray}
&&\frac{D^2\eta^{\hat{1}}}{D\tau ^2}=
\frac{1}{2} \left(\frac{2 \gamma ^2 \left(-\frac{2}{\lambda }-1\right) r^{-\frac{2}{\lambda }-2}}{\lambda }+\frac{2 R_s}{r^3}\right) \eta^{\hat{1}}\,,\\\label{radial1lambda}
&&\frac{D^2\eta^{\hat{i}}}{D\tau ^2}= \frac{1}{2r}\left(\frac{2 \gamma ^2 r^{-\frac{2}{\lambda }-1}}{\lambda }-\frac{R_s}{r^2}\right)\eta^{\hat{i}}\,.\label{angularilambda}
\end{eqnarray}
We plot the behaviour of the force components in Fig.\,\ref{Fig10}.
For the assumed values of $\lambda$ and $\Upsilon$, the radial tidal force has its maximum and minimum displaced away from $r=0$ when $\lambda$ and $\Upsilon$ $<0.7$. This implies that for $\lambda<1$ and $\Upsilon<1$ the radial component falls faster to $-\infty$ when compared to RN. There is a reversal in the radial elongation force for compression. This behaviour is similar to that previously obtained for \eqref{Ametric} with $l>0$. Note also that the behaviour analogous to RN is recovered in Fig.\,\ref{radlam1} (purple curve) and for Schwarzschild in Fig.\,\ref{radup1} (orange curve). The angular tidal force is zero at a point increasingly closer to $r=0$ as $\lambda$ and $\Upsilon$ increase, as can be seen in Fig.\,\ref{anglam1} and \ref{angup1}. There is initially compressive behavior due to angular forces, and as the object approaches a minimum force value, its elongation behavior reverses.  This behaviour was also observed for the metric  \eqref{Ametric}.  the behavior analogous to RN is recovered in Fig.\,\ref{anglam1} (purple curve) and for Schwarzschild in Fig.\,\ref{angup1} (orange curve).

\begin{figure*}[ht!]
\centering
\subfigure[$\Upsilon = 0.2$] 
{\label{CIradlam1}\includegraphics[width=8.5cm]{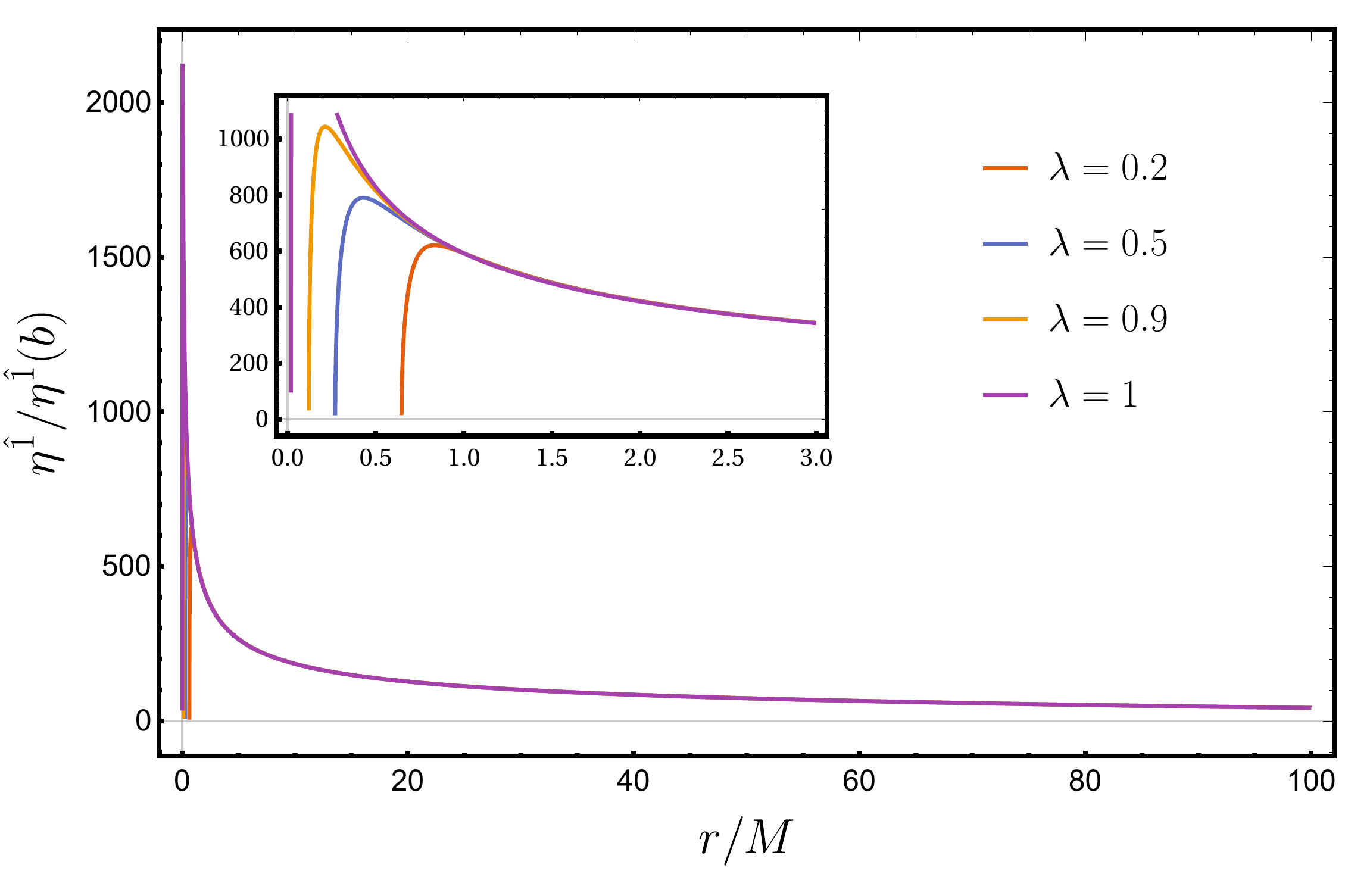} }
\hspace{0.1cm}
\subfigure[ $\lambda=0.6$] 
{\label{CIradup1}\includegraphics[width=8.5cm]{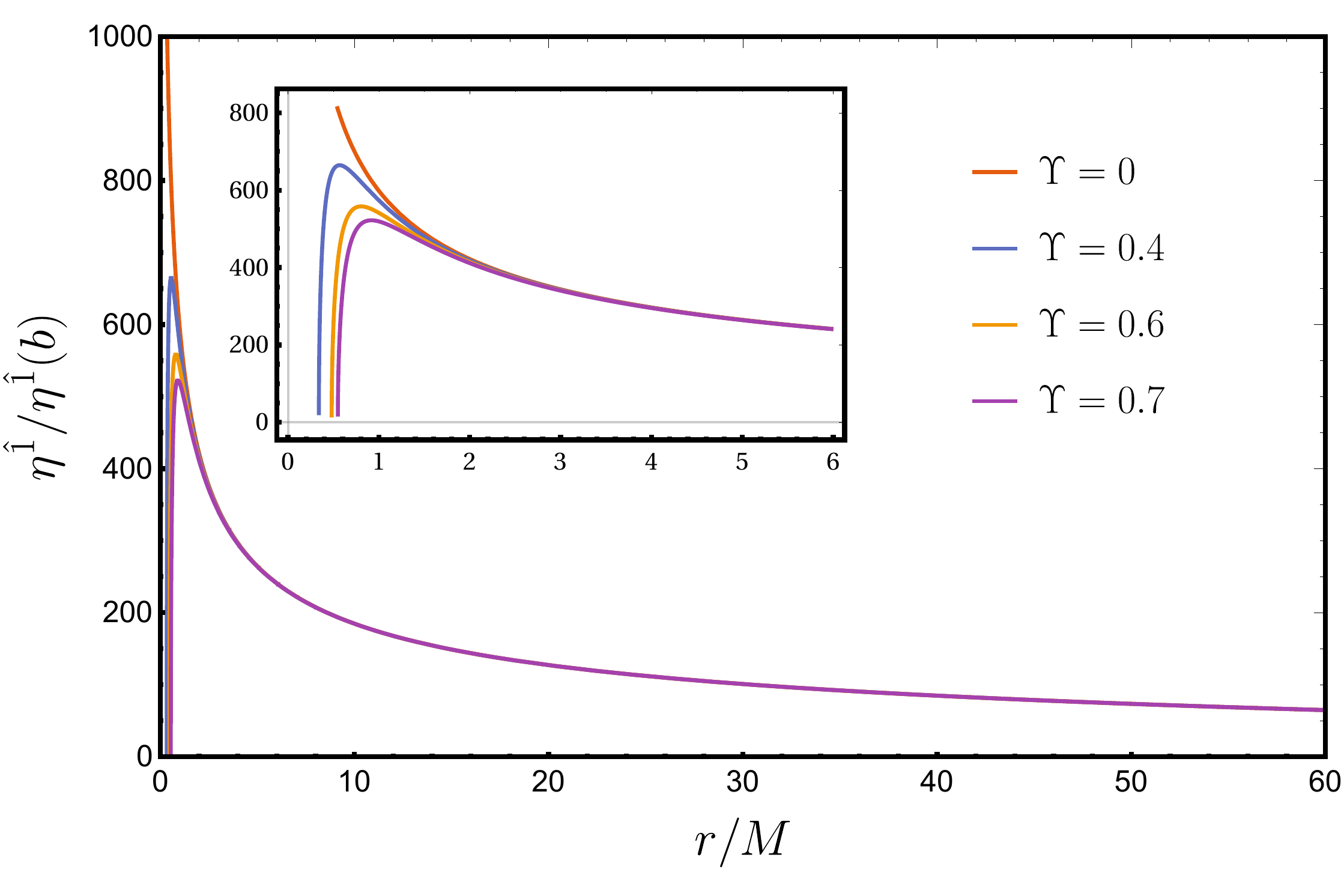} }
\caption{$\mathcal{CI}$: (a) Radial component of the geodesic deviation with $\Upsilon=0.2$ and different $\lambda$.  (b) Radial component of the geodesic deviation with $\lambda=0.6$ for different $\Upsilon$. 
}\label{Fig11}
\end{figure*}

The radial and angular components of the geodesic deviation for the solution \eqref{AKR2} are obtained, respectively, from the solution of the differential equations \eqref{eqdesv1} and \eqref{eqdesvi} using \eqref{Eb},
\begin{eqnarray}
\eta^{\hat{1}}(r)&&=\sqrt{j(r)}\Bigg[\frac{2\lambda b^{2+2/\lambda}}{b^{2/\lambda}R_s\lambda-2 b\Upsilon^2}\frac{d\eta^{\hat{1}}(b)}{d\tau}\nonumber\\
&&+\eta^{\hat{1}}(b)\frac{\left(\lambda R_s -2b^{1-2/\lambda}\Upsilon^2 \right)}{\lambda 2b^2}\int \frac{dr}{\left(j(r)\right)^{3/2}}\Bigg]\,,\label{radialRN}
\end{eqnarray}
\begin{eqnarray}
\eta^{\hat{i}}(r)=r\left[\frac{\eta^{\hat{i}}(b)}{b}-b\frac{d\eta^{\hat{i}}(b)}{d\tau}\int\frac{dr}{r^2\sqrt{j(r)}}\right]\,,\label{angRN}
\end{eqnarray}
where $j(r)=R_s/r-R_s/b+b^{-2/\lambda}\Upsilon^2-r^{-2/\lambda}\Upsilon^2$. 
For condition $\mathcal{CI}$, the components of the geodesic deviation are shown in Fig.\ref{Fig11}.
The effect of the spontaneous Lorentz symmetry breaking parameters is analogous to that found for the solution \eqref{Ametric}. However, in this case only positive values of this parameter are considered, as reported in \cite{KRlambda}. It is observed that the geodesic deviation reaches a limiting value and decreases as the object falls toward the origin and $\lambda$ and $\Upsilon$ increase. Again, this effect is analogous to that obtained for RN (see Fig.\,\ref{CIradlam1} for $\lambda=1$) and opposite to that obtained for Schwarzschild (see Fig.\,\ref{CIradup1} for $\Upsilon=0$), where the object crosses the event horizon and diverges to infinity due to the radial tidal force. Here, condition $\mathcal{CI}$ yields the identical result for the angular component \eqref{angRN} of the geodesic deviation as shown in Fig.\,\ref{Fig7}, since in this case $ \eta^{\hat{i}}(r)=r\frac{\eta^{\hat{i}}(b)}{b}$.

The second condition, $\mathcal{CII}$, provides the behaviour of the radial and angular components of the geodesic deviation for an object released near the black hole.
\begin{figure*}[ht!]
\centering
\subfigure[$\Upsilon = 0.2$] 
{\label{CIIradlam1}\includegraphics[width=8.5cm]{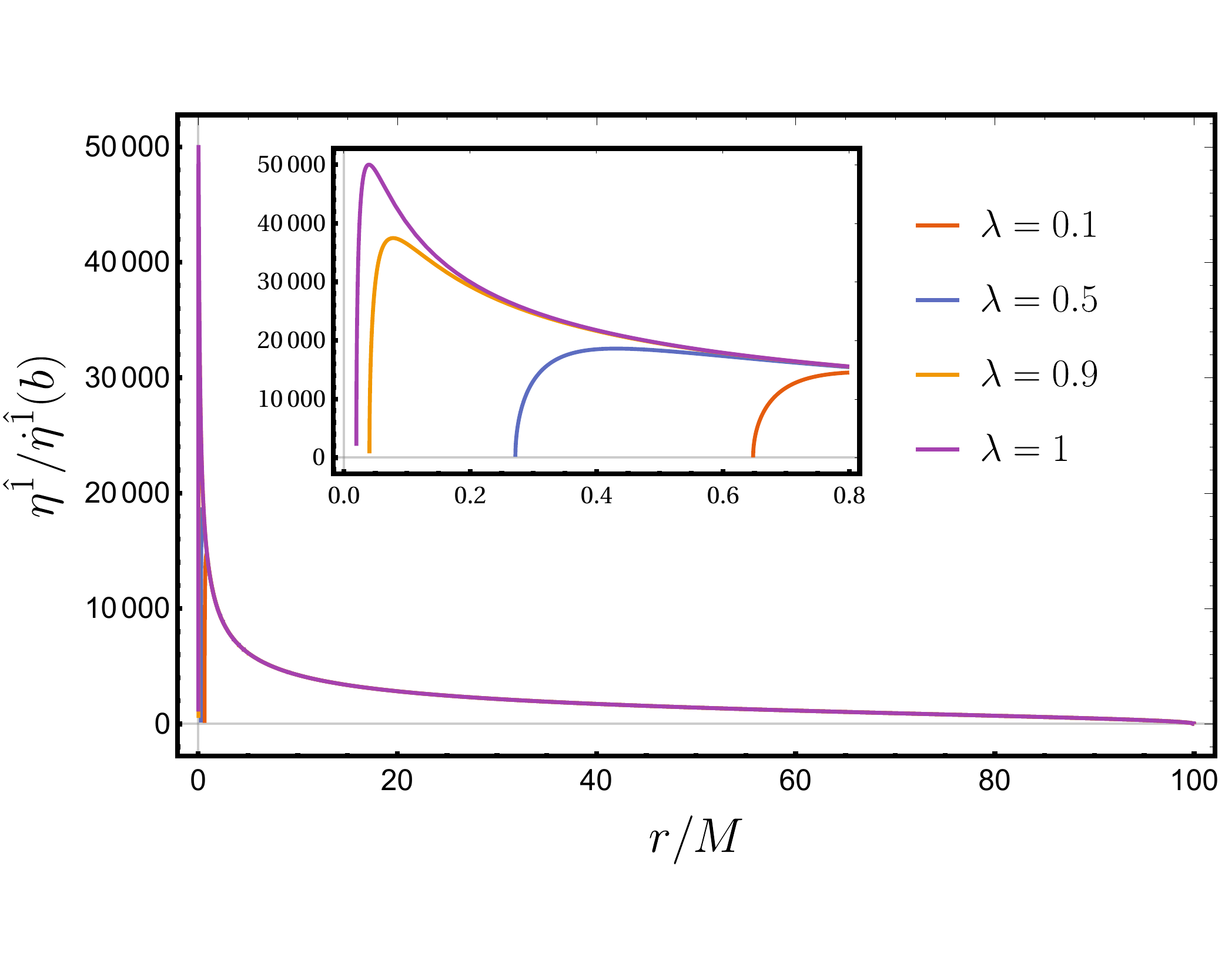} }
\hspace{0.1cm}
\subfigure[ $\lambda=0.6$] 
{\label{CIIradup1}\includegraphics[width=8.5cm]{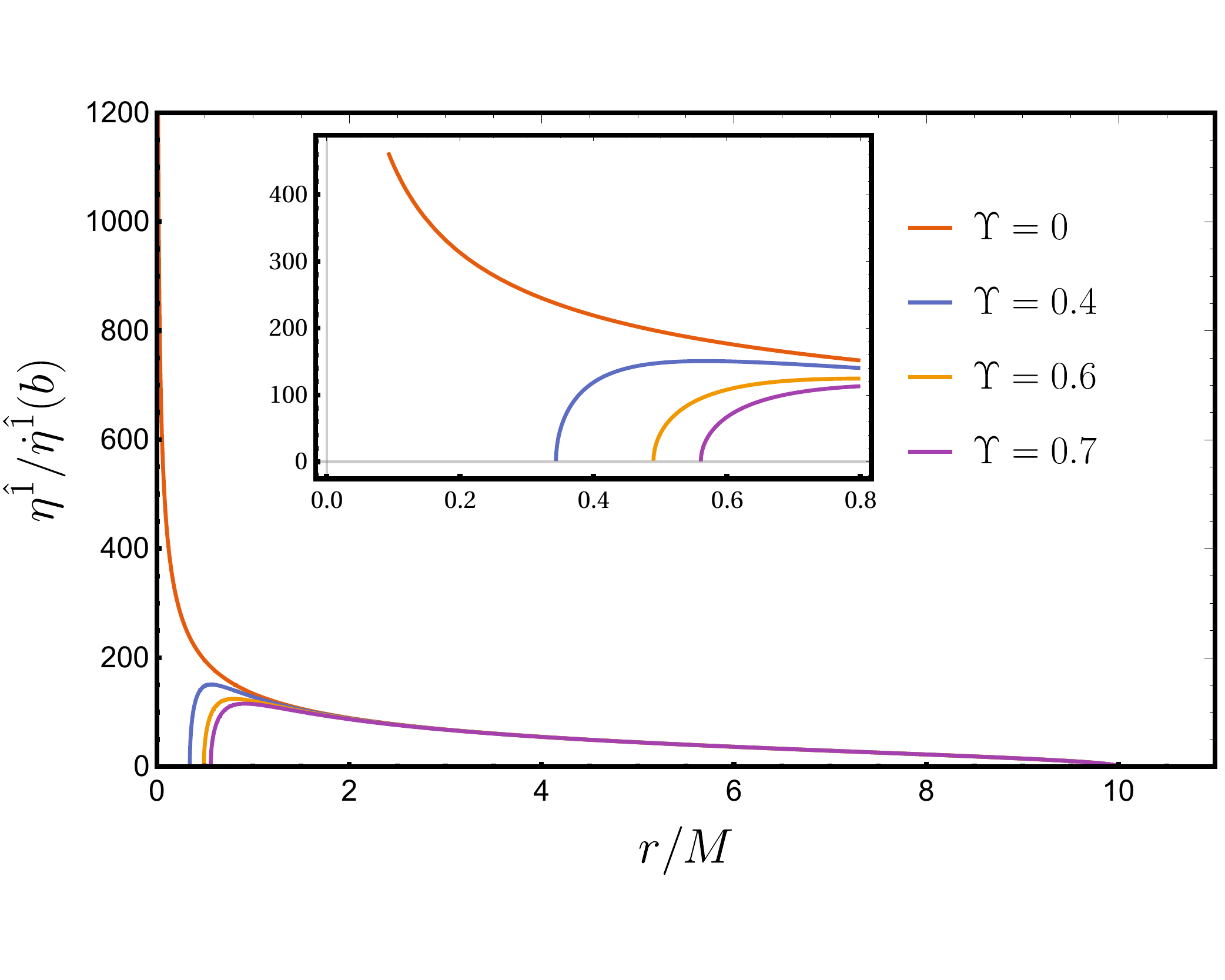} }
\caption{$\mathcal{CII}$: (a) Radial component of the geodesic deviation with $\Upsilon=0.2$ and different $\lambda$.  (b) Radial component of the geodesic deviation with $\lambda=0.6$ for different $\Upsilon$. 
}\label{Fig12}
\end{figure*}
Fig.\,\ref{Fig12} shows a behaviour similar to that of $\mathcal{CI}$ for the radial component of the geodesic deviation. The angular component behaves similarly to that obtained for \eqref{Ametric}, as seen in Fig.\,\ref{Fig13}, where the geodesic deviation initially increases up to a maximum at $b/2$ along the object's trajectory, and then begins to decrease due to the compression of the tidal forces. Here, the effect of the Lorentz symmetry breaking parameter is also subtle at large distances, being significant only near the origin, and differs from the Schwarzschild and RN cases. However, it resembles the behaviour observed previously, where after reaching a minimum, the angular component begins to increase up to a stopping point $r=R^{\rm stop}$, which varies as $\lambda$ and $\Upsilon$ change. Again, RN is recovered for $\lambda=1$ (Fig.\,\ref{CIIang1}) and Schwarzschild for $\Upsilon=0$ (Fig.\,\ref{CIIang2}).
\begin{figure*}[ht!]
\centering
\subfigure[$\Upsilon = 0.2$] 
{\label{CIIang1}\includegraphics[width=8.5cm]{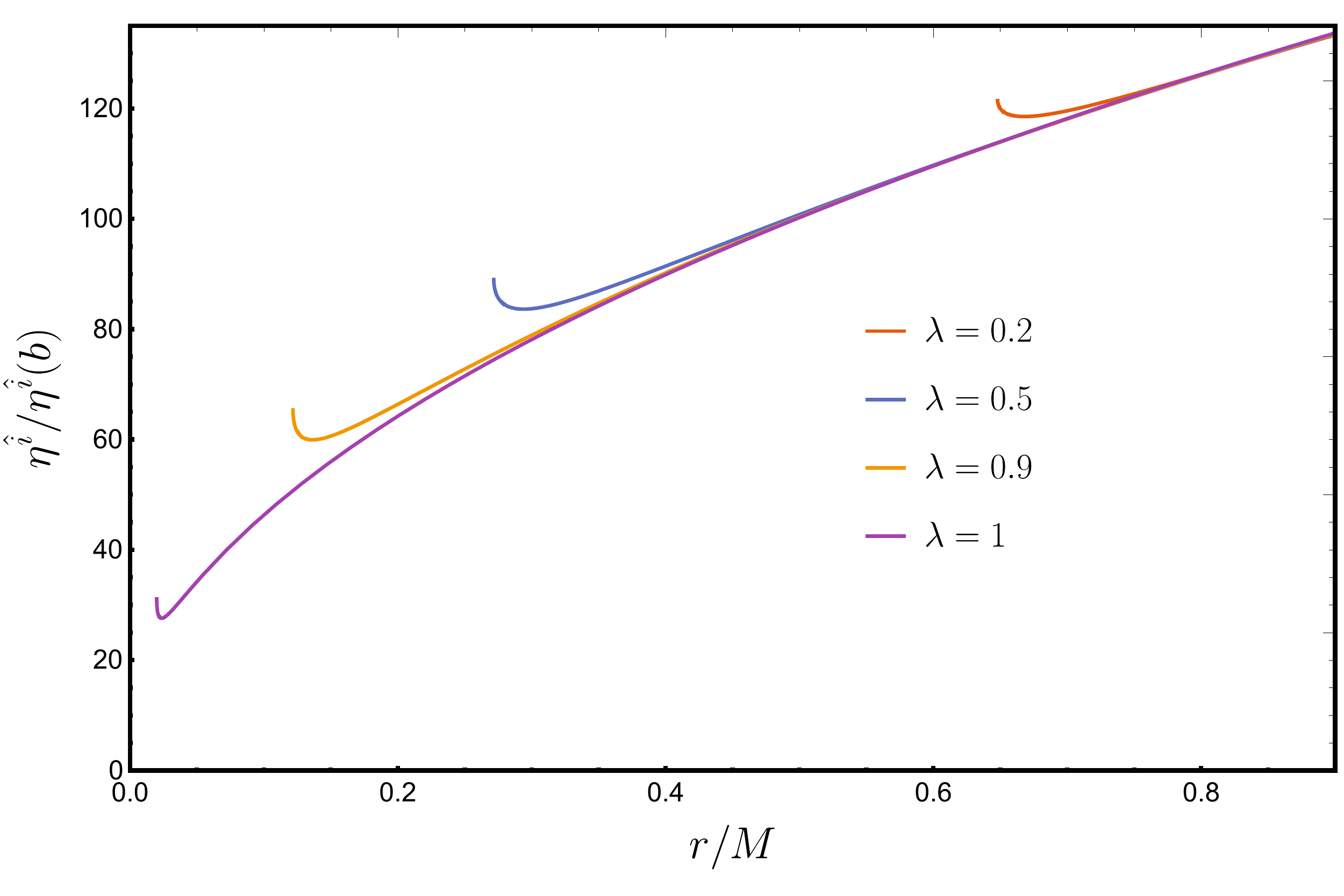} }
\hspace{0.1cm}
\subfigure[ $\lambda=0.6$] 
{\label{CIIang2}\includegraphics[width=8.5cm]{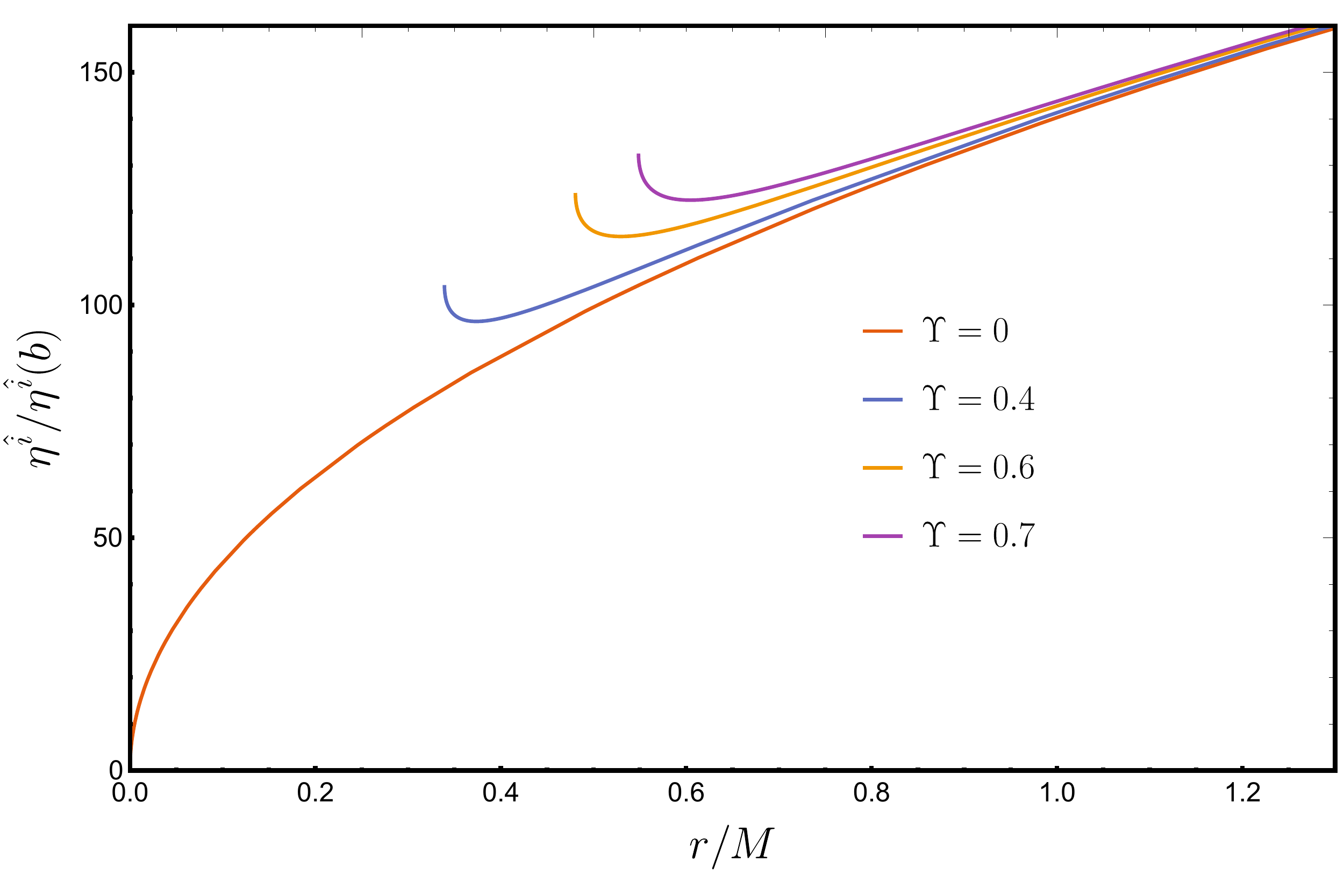} }
\caption{$\mathcal{CII}$: (a) Angular component of the geodesic deviation with $\Upsilon=0.2$ and different $\lambda$.  (b) Angular component of the geodesic deviation with $\lambda=0.6$ for different $\Upsilon$. 
}\label{Fig13}
\end{figure*}

\subsection{Lorentz symmetry-breaking in bumblebee model}

Now, the tidal effects will be evaluated for a metric obtained from a Schwarzschild-like solution in \cite{PetrovX}, which breaks Lorentz symmetry through a coefficient $X=\xi b^2$, within a metric-affine bumblebee gravity model:
\begin{eqnarray}
ds^2=-\frac{\left(1-\frac{2M}{r}\right)}{\kappa_1}dt^2+\frac{\kappa_2}{\left(1-\frac{2M}{r}\right)}dr^2+r^2d\Omega^2\,,\label{Petrovds}
\end{eqnarray}
where $\kappa_1=\sqrt{(1+3X/4)(1-X/4))}$  and $\kappa_2=\sqrt{(1+3X/4)/(1-X/4)^3}$ contains the Lorentz symmetry breaking coefficient.  The Schwarzschild metric is recovered in the limit $X \rightarrow 0$. The parameter $X$ was constrained in \cite{PetrovX} to $X < 10^{-12}$,  from the observational data of the advance of Mercury’s perihelion and for $X<10^{-10}$ using the very-long-baseline interferometry for light bending. These constraints yield numerical values very close to 1 for $\kappa_1$ and $\kappa_2$, such that the tidal effects, although present, are rather subtle. Considering this, we shall adopt values of $X<1$, as previously used in \cite{PetrovX2, PetrovX3}, in order to more clearly assess the role of Lorentz symmetry-breaking in the tidal effects for this space-time.
Since the symmetry $A(r)=1/B(r)$, where $A(r)$ and $B(r)$ are the metric functions of \eqref{Petrovds} in the form of \eqref{metricageral}, does not hold, it is necessary to generalize the equations for the force components and the geodesic deviation. We follow the procedure adopted in \cite{Marcostidal}. Therefore, equations\,\eqref{radial1} and \eqref{angulari} take the generalized form,
\begin{eqnarray}
&&\frac{D^2\eta^{\hat{1}}}{D\tau ^2}=\left(-\frac{B(r)A''(r)}{2A(r)}-\frac{A'(r)B'(r)}{4A(r)}+\frac{B(r)A'(r)^2}{4A(r)^2}\right)\eta^{\hat{1}}\,,\nonumber\\
\label{radial1Marcos}\\
&&\frac{D^2\eta^{\hat{i}}}{D\tau ^2}=\left(\frac{(E^2-A(r))B'(r)}{2rA(r)}-\frac{E^2B(r)A'(r)}{2rA(r)^2}\right)\eta^{\hat{i}}\,.\label{angulariMarcos}
\end{eqnarray}
Therefore, by using \eqref{Petrovds} in these equations, we obtain the radial and angular tidal forces.
\begin{eqnarray}
&&\frac{D^2\eta^{\hat{1}}}{D\tau ^2}=\frac{2M}{\kappa_2 r^3}\eta^{\hat{1}}\,,\label{F1Petrov}\\
&&\frac{D^2\eta^{\hat{i}}}{D\tau ^2}=-\frac{M}{\kappa_2 r^3}\eta^{\hat{i}}\,.\label{FiPetrov}
\end{eqnarray}

\begin{figure*}[ht!]
\centering
\subfigure[$b=100$.] 
{\label{Fradpetrov}\includegraphics[width=8.5cm]{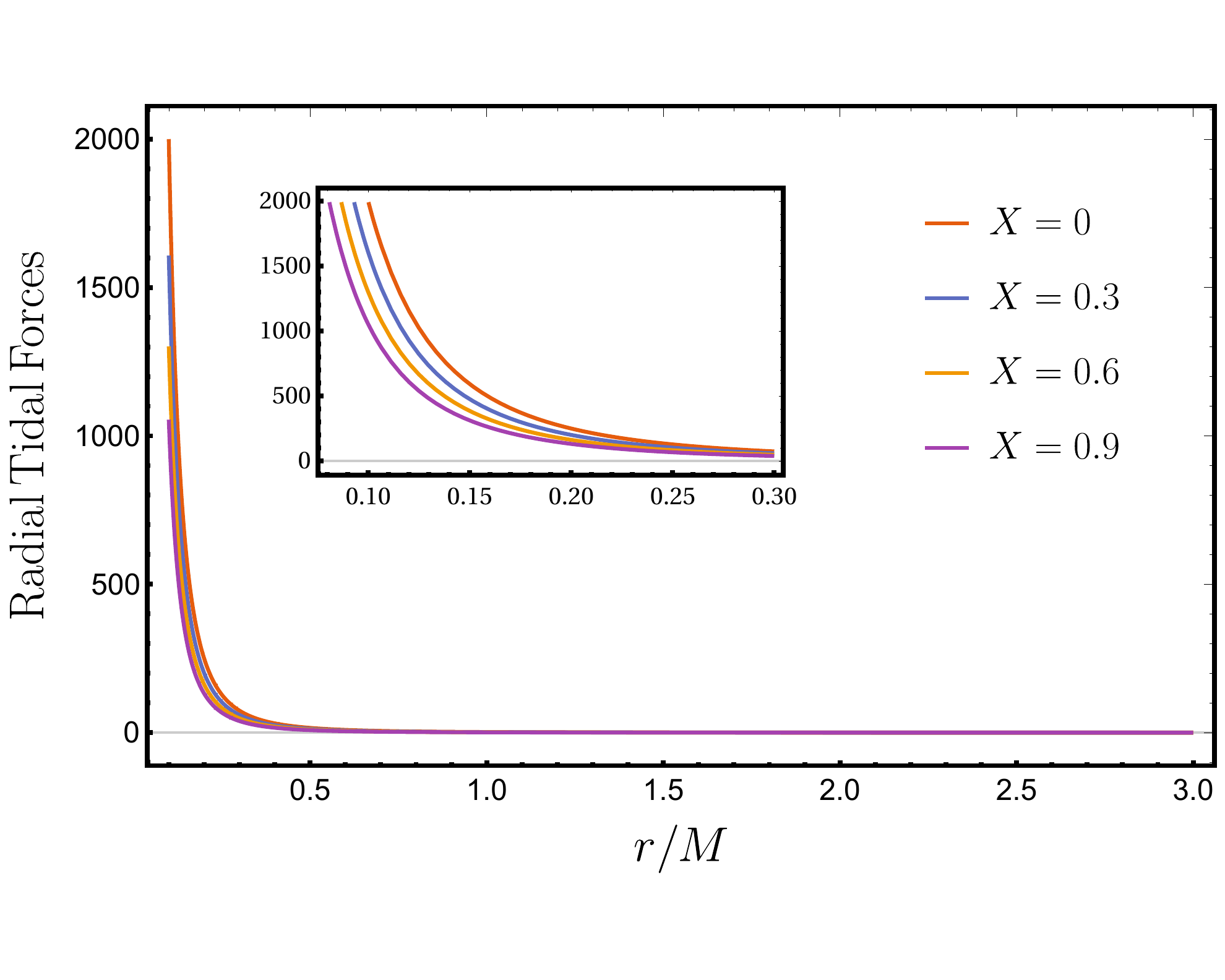} }
\hspace{0.1cm}
\subfigure[ $b=100$.] 
{\label{Fangpetrov}\includegraphics[width=8.5cm]{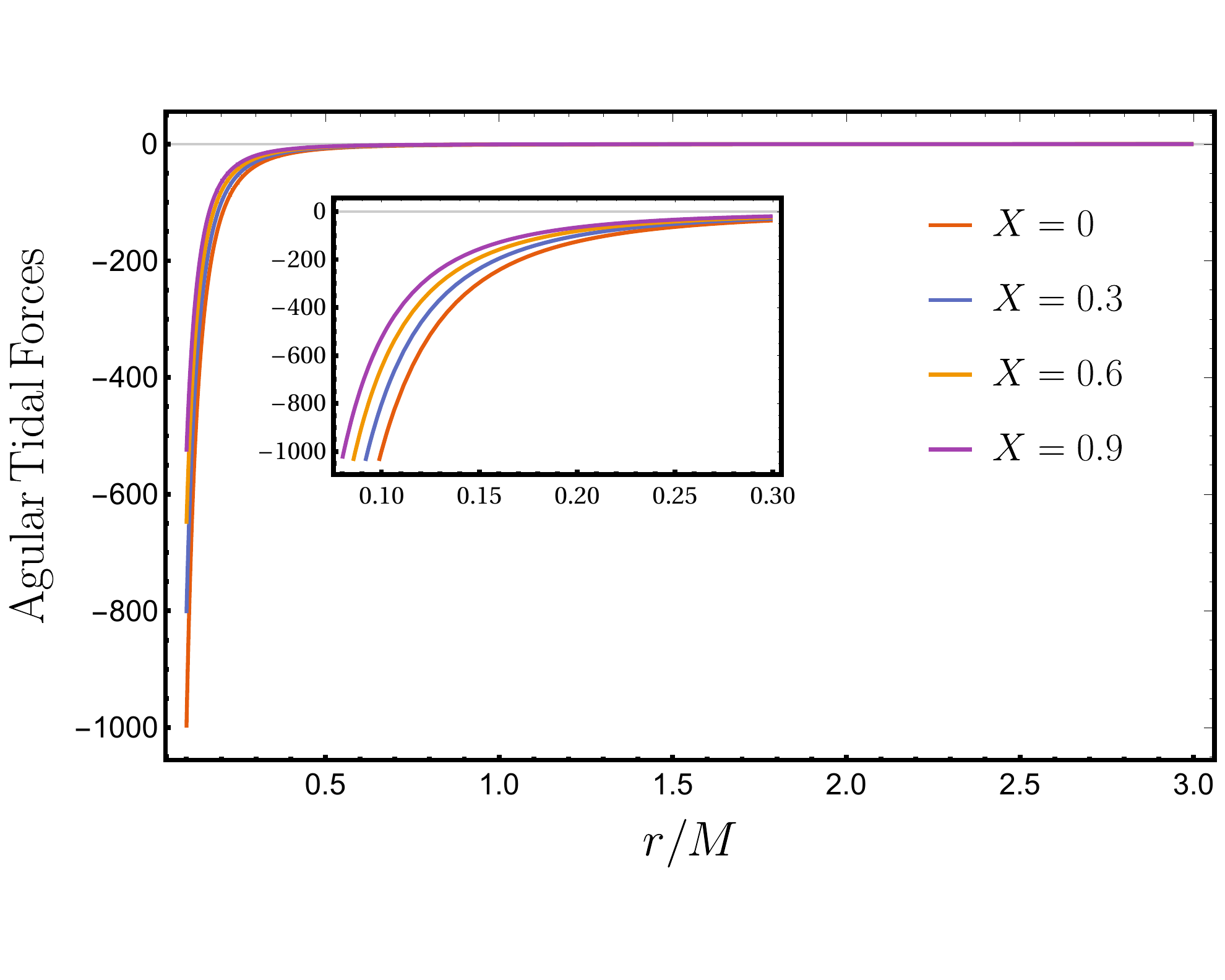} }
\caption{Radial tidal forces for different $X$.  (b) Angular tidal forces for different $X$.
}\label{Fig14}
\end{figure*}

Fig.\,\ref{Fig14} shows the influence of the Lorentz symmetry breaking parameter $X$ on the radial and angular components of the tidal forces. 
We note a clear difference in the tidal effects compared to the two space-times studied here, while showing similarity with the Schwarzschild case. For this solution, there are no compression regions for the radial component, only stretching, and no regions where the angular component is stretched, only compressed, as expected for Schwarzschild. However, as the Lorentz symmetry breaking parameter $X$ increases, the divergent behaviour of the tidal forces, in both radial and angular components, occurs more slowly as $r \rightarrow 0$. The tidal effects may still be small near the event horizon, but grow dramatically toward the singularity.

\begin{figure*}[ht!]
\centering
\subfigure[$b=100$.] 
{\label{RPtov}\includegraphics[width=8.5cm]{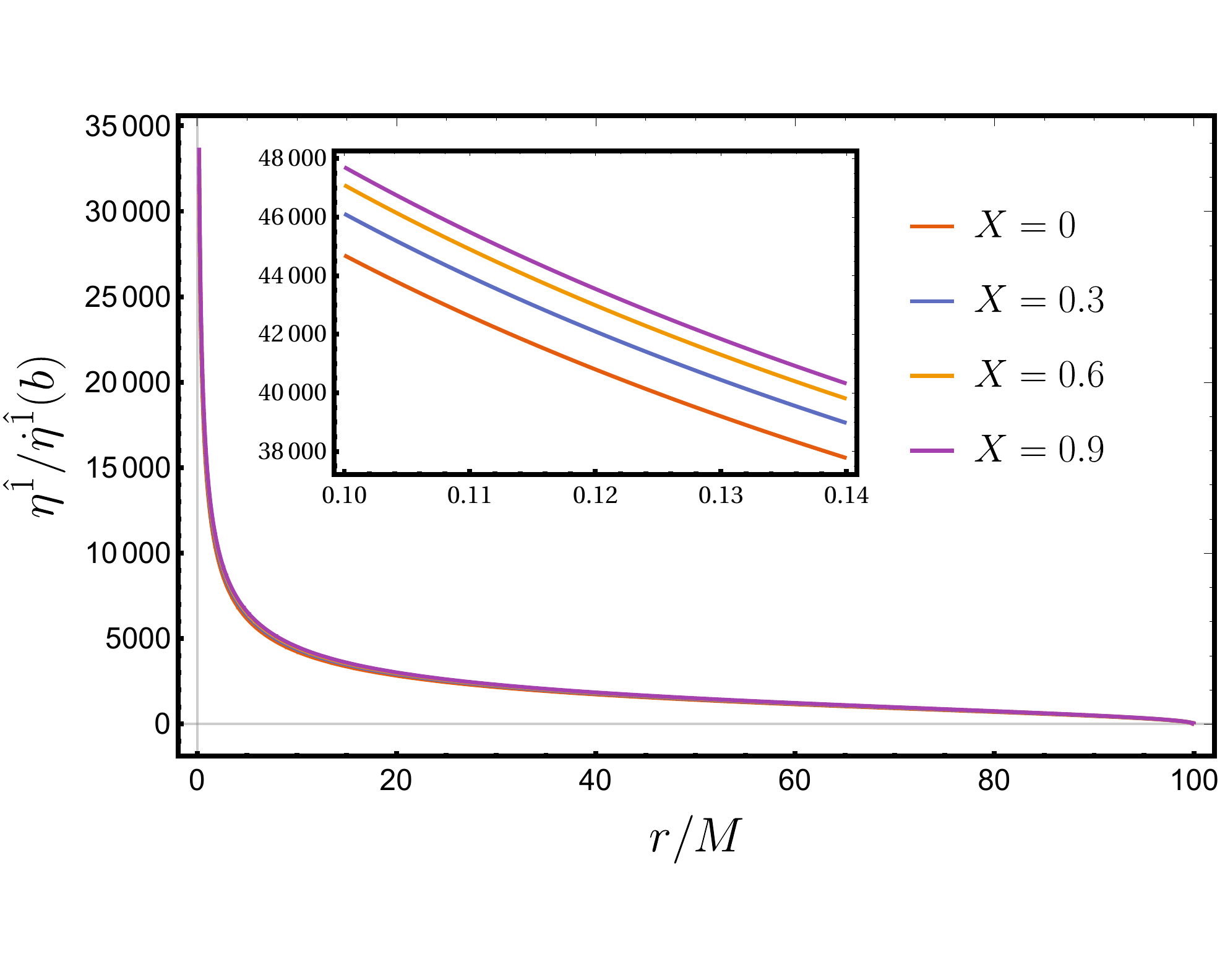} }
\hspace{0.1cm}
\subfigure[ $X=0.6$.] 
{\label{ANtrov}\includegraphics[width=8.5cm]{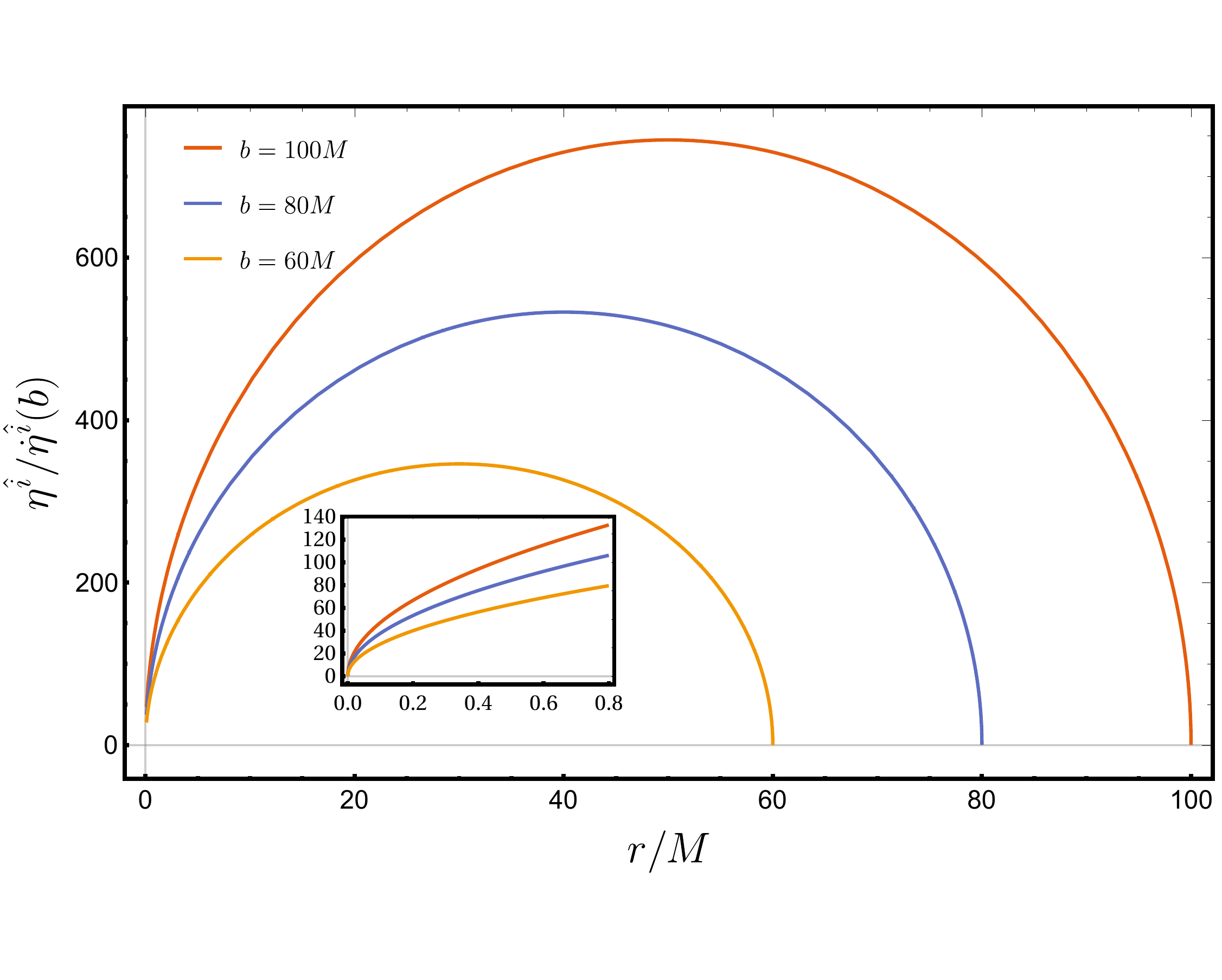} }
\subfigure[$b = 100$.] 
{\label{ANVX}\includegraphics[width=8.5cm]{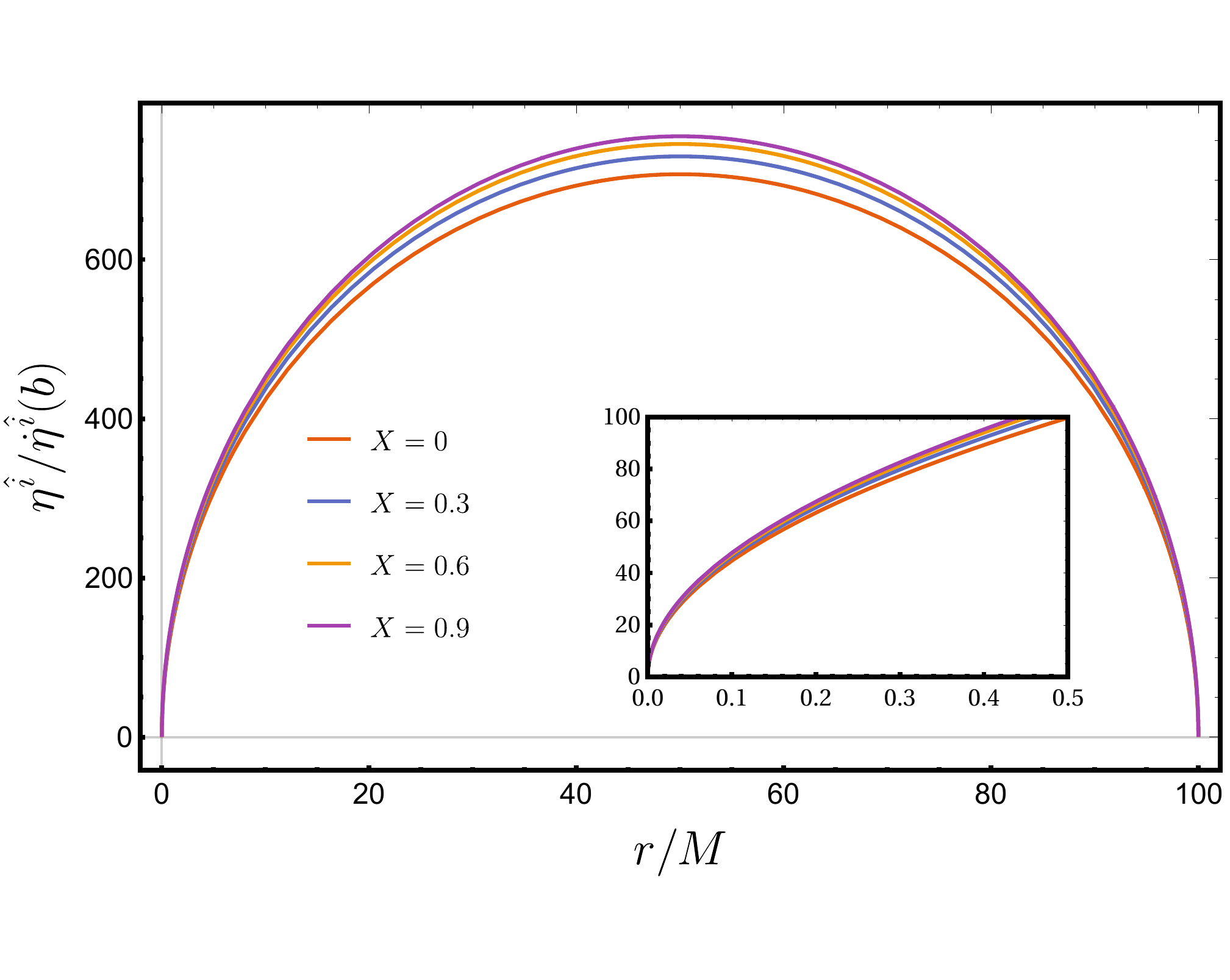}}
\caption{$\mathcal{CII}$: (a) Radial component of the geodesic deviation with  different $X$.  (b) Angular component of the geodesic deviation  for different $b$ and $X=0.6$.  (c) Angular component of the geodesic deviation  for different $X$.
}\label{Fig15}
\end{figure*}

We now proceed to evaluate the components of the geodesic deviation. To this end, we rewrite \eqref{rponto} in a generalized form, 
\begin{eqnarray}
-\frac{E^2}{A(r)}+\frac{\dot{r}^2}{B(r)}=-1\,,
\end{eqnarray}
and use it to convert Eqs.\,\eqref{radial1Marcos} and \eqref{angulariMarcos} into differential equations in $r$, taking the metric \eqref{Petrovds}, 
\begin{eqnarray}
r \Big[r \eta^{\hat{1}}(r)'' \left(E^2 \kappa_1 r+2 M-r\right)-&&M \eta^{\hat{1}}(r)'\Big]\nonumber\\
&&-2 M \eta^{\hat{1}}(r)=0\,,\nonumber
\end{eqnarray}
\begin{eqnarray}
 \left[r \eta^{\hat{i}}(r)'' \left(E^2 \kappa_1 r+2 M-r\right)-M \eta^{\hat{i}}(r)'\right]+M \eta^{\hat{i}}(r)=0\,,\nonumber
\end{eqnarray}
which we then solve to obtain the following radial and angular components:
\begin{eqnarray}
\eta^{\hat{1}}(r)=&&\sqrt{\frac{2b}{r}\frac{M(b-r)}{\kappa_1}}\frac{\kappa_1}{M}\frac{d\eta^{\hat{1}}(b)}{d\tau}+\frac{1}{2}\eta^{\hat{1}}(b)\left(3-\frac{r}{b}\right)\nonumber\\
&&+\frac{3}{2}\sqrt{\frac{b}{r}-1}\,\arccos\left[\left(\frac{r}{b}\right)^{1/2}\right]\eta^{\hat{1}}(b)\,,\label{desvPetrovrad}
\end{eqnarray}
\begin{eqnarray}
&&\eta^{\hat{i}}(r)=\left[\frac{1}{b}\eta^{\hat{i}}(b)+\kappa_1\frac{d\eta^{\hat{i}}(b)}{d\tau}\sqrt{\frac{2b}{M}\left(\frac{b}{r}-1\right)}\right]r\,.\label{desvPetrovang}
\end{eqnarray}
It is easy to see that when $X \rightarrow 0$, $\kappa_1 \rightarrow 1$ and the Schwarzschild results are fully recovered. Fig.\,\ref{Fig15} shows the behaviour of the radial and angular geodesic deviation. Under condition $\mathcal{CI}$, the components \eqref{desvPetrovrad} and \eqref{desvPetrovang} reduce to the usual Schwarzschild case, since $\frac{d\eta^{\hat{1}}(b)}{d\tau}=\frac{d\eta^{\hat{i}}(b)}{d\tau}=0$.

These results have already been thoroughly discussed in this manuscript as the limiting case of the other solutions, and are therefore not plotted here. However, the influence of the Lorentz symmetry-breaking parameter can be observed under condition $\mathcal{CII}$. For the radial component of the geodesic deviation, shown in Fig.\,\ref{RPtov}, the behavior is identical to that of Schwarzschild, except that as $X$ increases, the object is stretched more slowly as it approaches the singularity, as expected.
In Figs.\,\ref{ANtrov} and \ref{ANVX}, similar to the previous cases, the angular components of the geodesic deviation increase up to a certain point, $b/2$, after which they begin to decrease due to the compressive effect of the angular forces. However, no sudden increase in the deviation is observed as the object approaches the origin, since there is no sign change in the angular force component, as occurs in the previous cases. Here, the effect is again identical to Schwarzschild, and $X$ acts to increase the geodesic deviation as its value grows.

\section{Summary and  discussion}\label{Sec:Conclusion}

In this work, we have considered a static, spherically symmetric solution incorporating the Kalb-Ramond (KR) field, which is further modified by the presence of a spontaneous Lorentz symmetry-breaking parameter \cite{KReletrico}. The gravitational background is characterized by three fundamental parameters: the mass $M$ of the black hole, the electric charge $q$, and the KR parameter $l$, which encodes deviations from the standard Reissner-Nordström solution due to the influence of the Lorentz-violating effects. 
Our primary objective was to investigate the free-fall motion of a test particle in the gravitational field of a charged KR black hole, analyzing the dynamics from the perspective of different reference frames.

We have shown that the presence of inner and outer horizons, along with the spontaneous Lorentz symmetry-breaking parameter, significantly alters the motion toward the black hole's central singularity. An observer at $b > r_+$ measures the velocity of the falling particle increasing towards the speed of light as it approaches the outer horizon. This behaviour is similar to Schwarzschild and RN black holes; however, the initial velocity depends on the charge $q$ and is modified by the parameter $l$ due to the transformation $E^2 \rightarrow 1/(1-l)$ as $r \rightarrow \infty$. 
Inside the horizon, a fixed observer records a decrease in velocity to a minimum at $r_{\rm min}$, which also depends on $q$ and $l$, followed by an increase up to $c$ at the inner horizon $r_-$. Upon crossing the event horizon $r_+$, an inversion of temporal and spatial coordinates occurs due to the timelike Killing vector becoming space-like inside the horizon and, consequently, the measured velocities are inverted. This behaviour, observed in both RN and Schwarzschild space-times (see Fig.\,\ref{Fig1}), was previously shown in \cite{queda1} for the Schwarzschild space-time, where the velocity decreases to zero at the final singularity after horizon crossing.
 
Furthermore, we calculated the frequency ratios between an electromagnetic signal emitted by a fixed observer and received by a free-falling observer moving toward the charged KR black hole. The results indicate that, similarly to the previous case, the charge $q$ and the parameter $l$ significantly influence the outcome. 
A signal emitted by the fixed observer and detected by the falling observer, moving toward the singularity, exhibits a redshift. As the falling observer approaches the outer horizon, the frequency ratio asymptotically approaches a fixed value of $1/2$, independent of $q$ or $l$. This effect can serve as an indicator of horizon crossing. Inside the horizon, for a Schwarzschild metric, the ratio continues decreasing to zero, as shown in \cite{queda1}. 
For the charged KR black hole, Fig.\,\ref{Fig2} illustrated a decrease in the frequency ratio due to space-time expansion, where a Doppler redshift is observed until reaching a characteristic minimum determined by $q$ and $l$. This is followed by a space-time contraction, during which a Doppler blueshift occurs. These effects also reflect the velocity behaviour of the test particle.

We also analyzed the effects of tidal forces using the general formalism for a spherically symmetric space-time in order to compute the radial and angular components of the tidal force for KR black holes. We found that the radial and longitudinal tidal forces remain identical to the Schwarzschild case, indicating no influence from the spontaneous Lorentz symmetry-breaking parameter.
In Fig.\,\ref{Fig3} we plotted the behaviour of the radial tidal force, showing its exclusive dependence on $l$ and $q$. The radial tidal force $R^{\rm rad}_0$ cancels out between the horizons. As seen in Fig.\,\ref{Fradiala}, there is a maximum force at $R^{\rm rad}_{\rm \max}$ for $q \neq 0$ and varying values of $l$, indicating a transition from radial elongation to compression as the object approaches the central singularity. Fixing $l = 1.24 \times 10^{-1}$, we plotted the radial force for different values of $q$ in Fig.\,\ref{Fradialb}. As $q$ increases, the tidal force shifts from radial elongation to compression. This behaviour was predicted for RN in \cite{Crispino} and contradicts the behaviour observed for Schwarzschild black holes.
The angular tidal force behaviour is shown in Fig.\,\ref{Fangulara}. It reaches a minimum at $R^{\rm ang}_{\rm min}$, where the force reverses direction towards elongation and cancels out at $R^{\rm ang}_0$ between the horizons. The parameter $l$ moves these points closer ($l<0$) or farther apart ($l>0$). In Fig.\,\ref{Fangularb}, we fix $l$ and vary $q$, and observe a similar behaviour as the one for RN, with the $l$ parameter  accelerating or delaying the radial and angular tidal forces.

We also analytically solved the geodesic deviation equation for an object in free fall towards the central singularity, examining the influence of $q$ and $l$  under two sets of conditions. The behaviour of the geodesic deviation vectors is similar to that found in \cite{Crispino} for RN. In Figs.\,\ref{Figdesva3} and \ref{Figdesvb3} we observe from the angular component of the geodesic deviation the compression of the tidal forces, with the effect of the parameter $l$  becoming significant only near the origin.

Finally in Sec.\ref{sec6}, we have shown that the tidal effects in RN-like solutions with Lorentz symmetry-breaking, \eqref{AKR2}, exhibit behavior analogous to that previously obtained for the modified metric \eqref{Ametric}, differing from Schwarzschild and RN only in regions close to the origin. We observe a displacement of the extrema of the radial force, reversals between stretching and compression, and a zero point of the angular force that approaches $r=0$ as the violation parameters increase. The geodesic deviation remains bounded, in contrast to the Schwarzschild case where it diverges at the horizon. For the bumblebee model metric \eqref{Petrovds}, the effects are essentially equivalent to Schwarzschild but softened by the Lorentz-violating parameter, which delays the divergent growth of tidal forces and geodesic deviation near the singularity.

The results found in this paper support that the free-fall motion of test particles can serve as an useful tool for investigating the effects of spontaneous Lorentz symmetry-breaking on the dynamics of objects in the vicinity of charged KR black holes. By analyzing the Doppler shifts, tidal forces, and geodesic deviation, we observed how the $l$ parameter influences the motion of test particles near the event horizon. These modifications provide valuable insight into the nature of the spacetime and the underlying geometry of the black hole,  and can be used to differentiate black holes exhibiting spontaneous Lorentz symmetry-breaking from the traditional Schwarzschild one. 


\section*{Acknowledgements}

MER thanks Conselho Nacional de Desenvolvimento Cient\'ifico e Tecnol\'ogico - CNPq, Brazil, for partial financial support. This study was financed in part by the Coordena\c{c}\~{a}o de Aperfei\c{c}oamento de Pessoal de N\'{i}vel Superior - Brasil (CAPES) - Finance Code 001.
FSNL acknowledges support from the Funda\c{c}\~{a}o para a Ci\^{e}ncia e a Tecnologia (FCT) Scientific Employment Stimulus contract with reference CEECINST/00032/2018, and funding through the research grants UIDB/04434/2020, UIDP/04434/2020 and PTDC/FIS-AST/0054/2021.
DRG is supported by the Agencia Estatal de Investigación Grant Nos. PID2022-138607NB-I00 and CNS2024-154444, funded by MICIU/AEI/10.13039/501100011033 (Spain).



\end{document}